\documentclass[twocolumn, iop]{openjournal}
\usepackage{float}
\usepackage{amsmath}

\usepackage[caption=false]{subfig}
\usepackage{hyperref} 
\usepackage{makecell}
\usepackage{array}
\usepackage{dsfont}
\usepackage{gensymb}
\usepackage{longtable}
\hypersetup{
    colorlinks=true,
    linkcolor=blue,
    filecolor=magenta,      
    urlcolor=cyan,
    pdftitle={Overleaf Example},
    pdfpagemode=FullScreen,
    }  
\usepackage{orcidlink}
\DeclareRobustCommand{\VAN}[3]{#2}
\let\VANthebibliography\thebibliography
\def\thebibliography{\DeclareRobustCommand{\VAN}[3]{##3}\VANthebibliography}

\shorttitle{Fast Rotators}
\shortauthors{Kamai \& Perets}

\begin{document}

\title{Too fast to be single:  Tidal evolution and photometric identification of stellar and planetary companions}

\author{Ilay Kamai \orcidlink{0009-0008-5080-496X}}
\thanks{E-mail: \href{mailto:ilay.kamai@campus.technion.ac.il}{ilay.kamai@campus.technion.ac.il}}
\affiliation{Physics Department, Technion: Israel Institute of Technology,
Haifa 32000,
Israel}

\author{Hagai B. Perets \orcidlink{0000-0002-5004-199X}}
\affiliation{Physics Department, Technion: Israel Institute of Technology,
Haifa 32000,
Israel}
\affiliation{ARCO, Open University of Israel, R'anana, 
Israel}

\begin{abstract}
Many stars, including those in binary or multiple systems, exhibit modified rotational evolution due to tidal interactions. While magnetic braking slows rotation in single stars, close binaries experience synchronization from tidal forces, resulting in high spin rates. Thus, fast rotators often signify synchronized binaries or planetary systems.
We analyze stellar rotation in the Kepler field to photometrically identify non-single systems. Establishing an initial rotation–temperature relationship for individual stars via young clusters, we confirm our findings through magnitude excess and prior binary star system studies. Stars rotating faster than this relationship display a bimodal distribution in peculiar velocity, indicative of non-single or young stars. Leveraging this, we separate non-single stars when peculiar velocity is measurable, or estimate likelihood for those without. Our method identifies 2229 potential non-single star systems with rotation periods exceeding 3 days. For ultra-fast rotators ($P_{rot} < 3 $ days), we compile a catalog of 1518 ultra-short-period binary candidates, often part of hierarchical triples, reinforcing rapid spin's association with multiplicity.
Applying our method to planet-host stars uncovers Kepler-1184 as a potential circumbinary system and identifies Kepler-493 and Kepler-957 potentially synchronized by close-in planets, with three others as potential false positives. Analysis of known non-single stars reveals clear tidal effects: period synchronization, orbit circularization, and a minimal pericenter constraint for binaries ($r_p \propto (P_{orb}/P_{rot})^{0.77}$).
These findings offer insights into tidal evolution, provide a robust method for identifying stellar multiplicity, and have implications for stellar evolution, binary formation, and exoplanet dynamics.
\end{abstract}

\keywords{astronomy  --- light curve --- binaries}

\section{Introduction} \label{sec:intro}
The stellar rotation of stars, their spins, can significantly affect their evolution, and the spins themselves are affected by various physical processes. The spin properties of stars can, therefore, be used as indicators for physical properties affecting their evolution and even the identification of interactions with other stellar or planetary companions. 

Single stars lose angular momentum throughout their lifetime. This is a result of the interaction between the stellar magnetic field and mass taken by stellar winds and causes a change in the angular momentum of the wind compared to that of the stellar surface. This, in turn, causes single stars to spin down over time, a process known as magnetic braking \citep{Schatzman1962, Weber1967, Mestel1968, Mestel1987}. This led to the use of the rotation period as a stellar age indicator \citep{Skumanich1972}, in a process called gyrochronology \citep{barnes2003, barnes2007, Mamajek2008, Angus2015, Angus2019, Bouma2023, Bouma2024}. However, in closely separated binaries, the situation is different. Although the total angular momentum is conserved in close binaries, to first approximation, those systems are subject to mutual tidal forces that distort their stellar shape, breaking their spherical and axial symmetry. The tidal torques gradually circularize and synchronize the stellar and orbital periods, preventing the spin-down of the components and even spinning them up. Here, we make use of the expected resulting fast-spinning stars produced in such interactions to identify such close binary interactions, and hence identify binary and other types of multiple systems through the photometrically measured spins of stars. 

The theory of tidal circularization and synchronization suggests that tidal torque can be decomposed into two different components:
\begin{itemize}
    \item \textit{Equilibrium Tides} - This force results from a non-wavelike, quasi-hydrostatic tidal bulge. The delay of the hydrostatic bulge is due to the coupling of the tidal flow to the motion of turbulent eddies in the stellar convective envelope. The coupling creates a phase shift between the tidal bulges and the orbital motion, which results in a torque between the two stars \citep{Zahn1977, Hut1981, Ogilvie2014}.
    \item \textit{Dynamical Tides} -  Describes the excitation and damping of gravity (g) waves in the radiative zones due to perturbations of the spherical mass distribution of the stars, caused by tidal potential \citep{Zahn1977, Savonije1983, Goodman1998, Ogilvie2014}. 
\end{itemize}
Equilibrium tides are effective in late-type stars with radiative cores and convective envelopes, and dynamical tides are more effective in early-type stars with radiative envelopes. However, this is not a distinct separation since dynamical tides have also been applied to the radiative cores of late-type stars \citep{Savonije2002}.  
Both theories predict a strong dependence of tidal synchronization on the semi-major axis \citep{Zahn1989, Claret1995}, which suggests that most short-period binaries are tidally locked. Indeed, synchronization was observed by \cite{Meibom2005}, \cite{Mazeh2008}, \cite{Van_Eylen2016}, and \cite{Lurie2017} which found that most binaries with $P_{orb} \leq 10$ days are synchronized. \cite{simonian2019} analyzed the magnitude displacement of a sample of \textit{Kepler} stars and concluded that fast rotators ($P_{rot} \leq 7$ Days) are dominated by synchronized binaries.
However, observational evidence of tidal interactions remains challenging to obtain, making each measurement particularly valuable.

In this work, we make use of two new datasets to further investigate tidal features and multiplicity - Gaia non-single stars catalog \citep{Gaia2023} and a catalog of main-sequence stellar rotation period by \cite{kamai2024}.

The paper is organized as follows: In section \ref{sec:data} we describe the data sample used in this study; in section \ref{sec:phot_binaries}, we use clusters to derive a cutoff temperature-period line for single stars. We validate this line with various methods, differentiating binaries from very young stars and calculating a general non-single probability for non-single candidates. In section \ref{sec:results} we apply this method to identify potential non-single stars (\ref{subsec:non_singles_catalog}) and triples (\ref{subsec:triples}); In  \ref{subsec:planets} we apply this method to planet host stars. in section \ref{sec:tidal_features}  we analyze tidal features in known non-single stars in the \textit{Kepler} field; In section \ref{sec:conclusions} we summarize our conclusions. 

\section{Data Sample} \label{sec:data}
In this study, we use the rotation periods from \cite{kamai2024}, which provides a catalog of 82771 \textit{Kepler} main sequence (MS) stars derived using a deep learning model called \textit{LightPred}. The fact that this catalog is the largest main sequence period catalog to date is due to the fundamental difference between data-driven models and classical algorithms (such as Autocorrelation function, Wavelet etc.). While the first would create predictions for every sample, the latter is usually constrained to only a sub-sample of valid points (for example, samples with Autocorrelation peaks above some threshold). The information on the 'validity' of samples in a data-driven model is manifested in the uncertainty of the model. For example, \cite{kamai2024} showed a strong correlation between their model's uncertainty and a 'periodicity' parameter defined in \cite{McQuillan2014}. For a detailed comparison between the different methods, please refer to the original paper. We slightly modified the final catalog from \cite{kamai2024} - in their paper, for each \textit{Kepler} sample, they took the average prediction of a moving window over the light curve. We took the median prediction to better account for outliers. We used a lower bound of $P_{rot} \geq 3$ days as \cite{kamai2024} found that below $3$ days their model is less reliable. The origin of this lower bound lies in the fact that their model was trained on simulated data, and many of the effects dominating very fast rotators (ellipsoidal variations, heart-beating pattern, etc.) were not modeled, making them an out-of-distribution samples. Finally, we remove potential giants according to the criterion given by \cite{Ciardi2011} - A star was considered to be a dwarf if the surface gravity was greater than the value specified by the following expression:
\begin{equation} \label{eq:giant_cond}
    log(g) = \begin{cases} 
    3.5 & \text{if } \mathrm{Teff} \geq 6000 \\
    4.0 & \text{if } \mathrm{Teff} \leq 4250 \\
    5.2 - (2.8 \times 10^{-4} \mathrm{Teff}) & \text{if } 4250 < \mathrm{Teff} < 6000
    \end{cases} 
\end{equation}
In their paper, \cite{kamai2024} performed a comprehensive error analysis and compared the results with previous results, such as the catalog from \cite{McQuillan2014}, and the Eclipsing Binaries (EB) catalog. They found that compared to the autocorrelation function, \textit{LightPred} model is better for fast rotators (above $3$ days) as it implicitly differentiates different stellar types, using its uncertainty level, and correctly reconstructs the observed tidal synchronization line. The autocorrelation function naively identifies the orbital period as the stellar period, resulting in an incorrect synchronization line. Moreover, the autocorrelation function cannot identify different stellar types (like ellipsoidal variations and pulsators), which can lead to wrong period predictions. This suggests that using the periods from \cite{kamai2024} should give a more accurate picture of fast rotators and the binary population. For a full description of the \textit{LightPred} model, please refer to the original paper. 

For other stellar parameters (effective temperature, radius, mass, metallicity), We used the catalog published in \cite{Berger_2020}. We also used K-band magnitude measurements from 2MASS \citep{Skrutskie_2006}. 

In addition, we used two catalogs of known binaries: a catalog of 2920 Eclipsing Binaries (EBs) in the Kepler field \citep{kirk2016}, and the Gaia non-single stars catalog \citep{Gaia2023} (Gaia-nss), which comprises more than $800000$ binaries in the Gaia field. The Gaia-nss catalog consists of non-single stars found by three different methods - astrometric, photometric, and spectroscopic. The ability to identify a star as non-single depends on the mass ratio of the binary, which translates to luminosity excess, and the binary separation (or orbital period). One would expect binaries with relatively high mass ratios to be more detectable than other types of binaries. Binary separation also affects the detectability; closer binaries are, in general, easier to detect, but there are caveats. For example, for an astrometric solution, one needs the sources to be separated, which leads to a lower bound on the orbital period that can be detected with this method. \cite{Sullivan_2025} showed that different indicators of binarity in Gaia, like RUWE and G-magnitude residual, show the expected dependencies with angular separation and binary magnitude ratio. This implies a Malmquist bias toward more luminous binaries for deeper samples. This bias is enhanced by the fact that Kepler samples are usually fainter than Gaia-nss samples (see Figure 1 in \cite{Gaia2023}). Although the Gaia-nss catalog comprises different solution types, only some provide the binary orbital period as part of the solution, which is crucial in this work. These are orbital solutions, eclipsing binaries, SB1, and SB2. After cross-matching those with Kepler data \footnote{we used cross match from \url{https://gaia-kepler.fun/} created by Megan Bedell with 1 arcsec radius}, we were left with 1988 samples.

\section{Photometric Identification of stellar multiplicity} \label{sec:phot_binaries}
We begin by investigating periods of clusters of stars at different ages. Since all the stars in each cluster are roughly the same age, we can distinguish age effects. Moreover, looking at relatively young clusters gives a good estimation for the initial period of stars on the main sequence. We use the clusters data that was used in \cite{Bouma2023} as calibration for a gyrochronology model. The data consists of $10$ different clusters with ages from $80$ Myr to $2.7$ Gyr from different surveys \citep{galindo2022, boyle2023, Curtis_2019, Rebull_2016, gillen2020, Curtis_2020, fritzewski_spectroscopic_2019, fritzewski_rotation_2021, messina_gyrochronological_2022, douglas_k2_2019, rampalli_three_2021, curtis_temporary_2019, jeffries_wocs_2013, meibom_spin-down_2015, torres_eclipsing_2020}. Besides rotation periods and effective temperature, \cite{Bouma2023} also flagged each star as a possible binary or single star. For a full description of the dataset, please refer to \cite{Bouma2023}. \\

\begin{figure*}
    \centering
    \includegraphics[width=\textwidth]{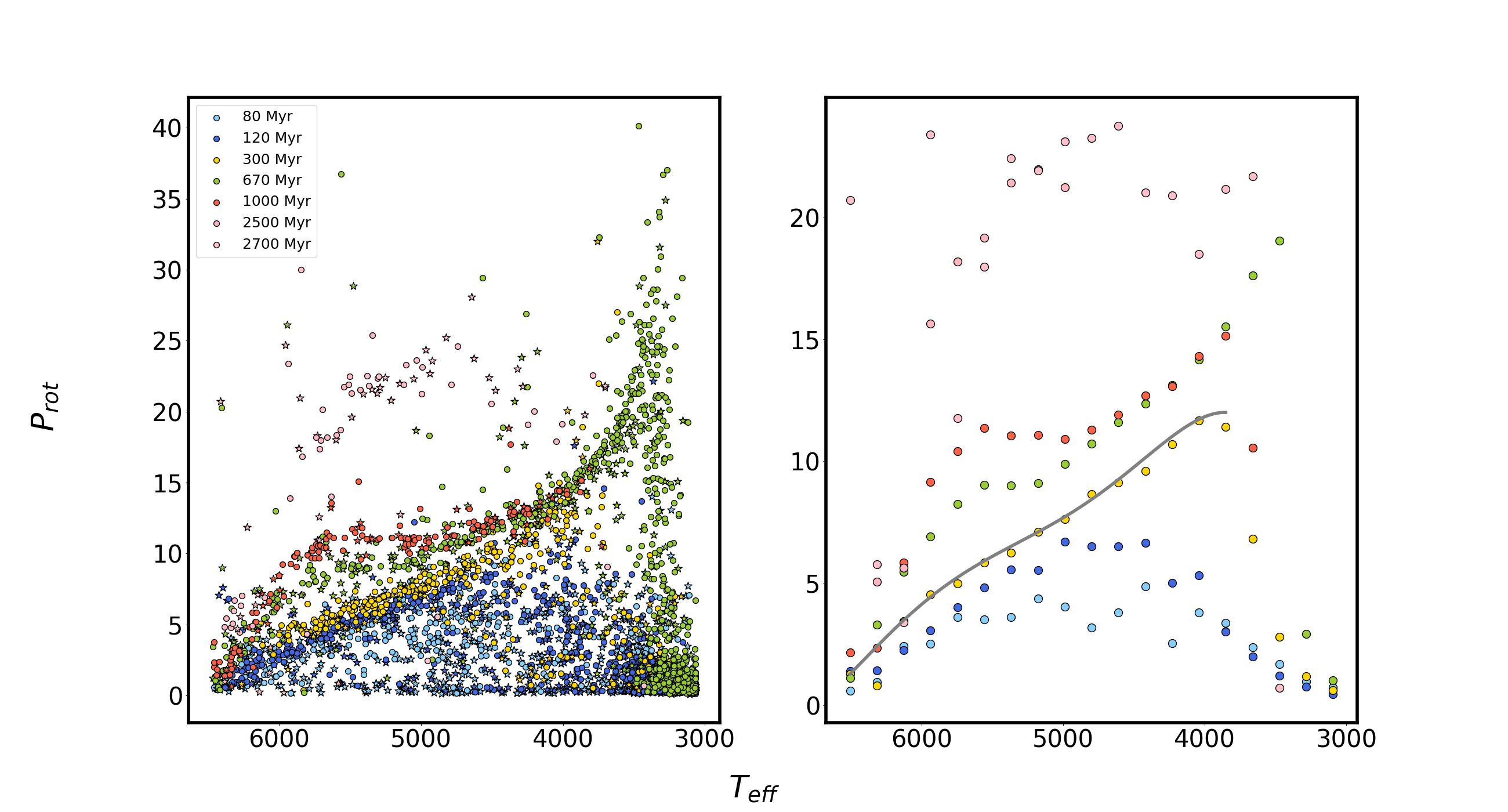}
    \caption{Left - Periods of young clusters as a function of temperature. Data is from \cite{Bouma2023}. Right - median values over $20$ temperature bins. The gray line is a $5^{th}$ order polynomial fitted to the gold data points for $3800 \leq T_{eff}\leq 6600$ } \label{fig:cluster_p}
\end{figure*}

Figure \ref{fig:cluster_p} shows the rotation periods vs. the effective temperature of the clusters data set. The left panel shows all the data points with colors representing the clusters' ages. Stars that were assigned as binary by \cite{Bouma2023} are marked as stars, and stars that were not assigned as binaries are marked as circles. One immediate observation is that hotter stars rotate faster and that young clusters ($\leq 300$ Myr) sit roughly on the same curve in the $T_{eff}-P_{rot}$ space. We also see that at cooler temperatures, there is more scatter, and the 'departure' from the curve is age-dependent: young clusters show a large scatter at hotter temperatures compared to older clusters. This is better seen in the right panel, which shows the median period over $20$ temperature bins for stars that were not assigned as binaries. We see that for young clusters, there is an 'inflection temperature' where the period-temperature relationships change. For temperatures lower than the inflection temperature, the period is either constant or decreases with the decrease of the temperature. We assign this change as the transition from the pre-main sequence to the main sequence. We want to construct a lower bound on the main-sequence period as a function of the temperature. As we already mentioned, the main sequence curves of clusters with age $\leq 300$ Myr are very similar. Since the $300$ Myr cluster is less scattered, we can take the curve resulting from the $300$ Myr cluster as a representative for the lower bound period. 

\begin{figure*}
    \centering
    \begin{minipage}[b]{0.45\textwidth}
    \includegraphics[width=\textwidth]{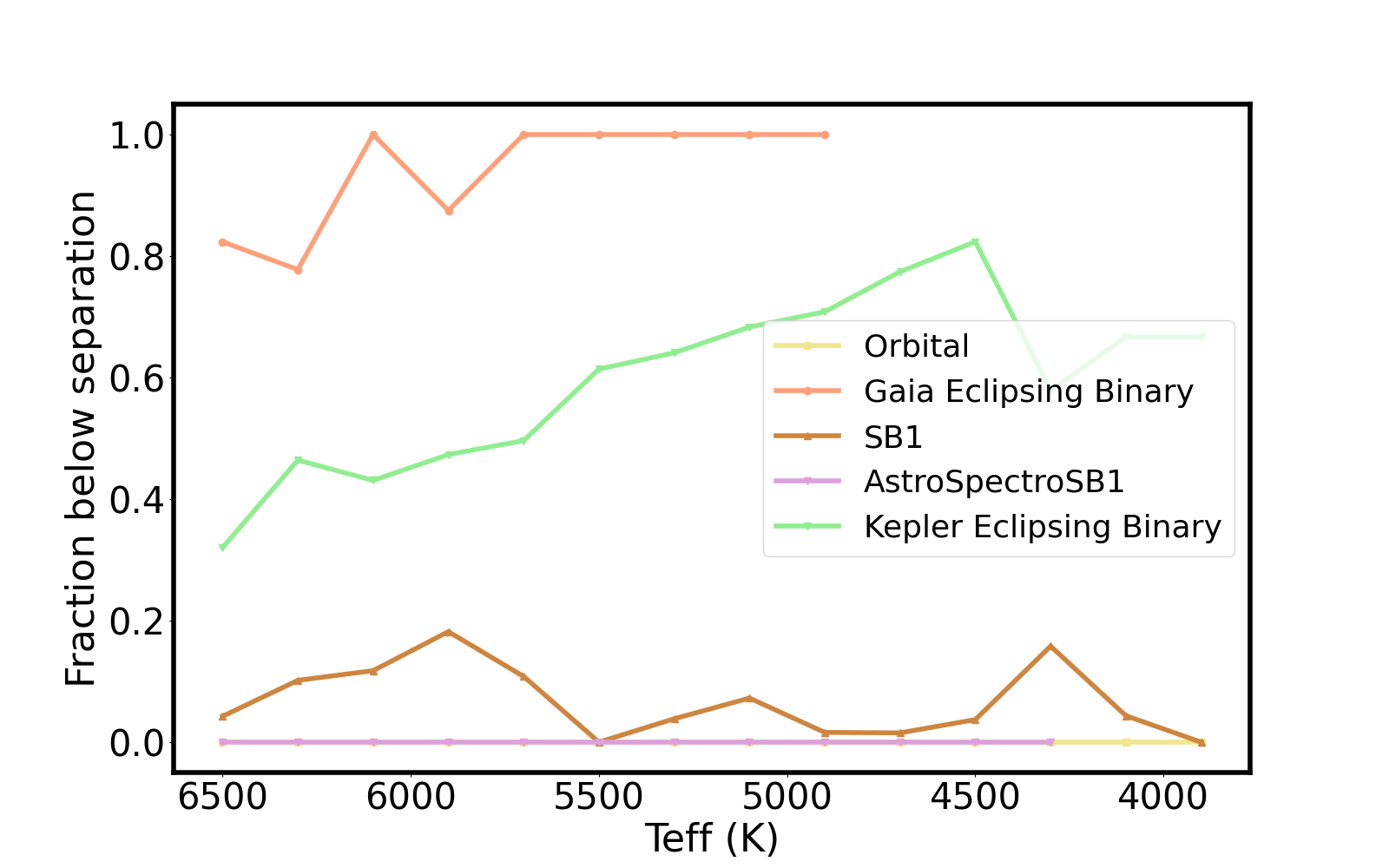}  
    \end{minipage}
    \begin{minipage}[b]{0.45\textwidth}
    \includegraphics[width=\textwidth]{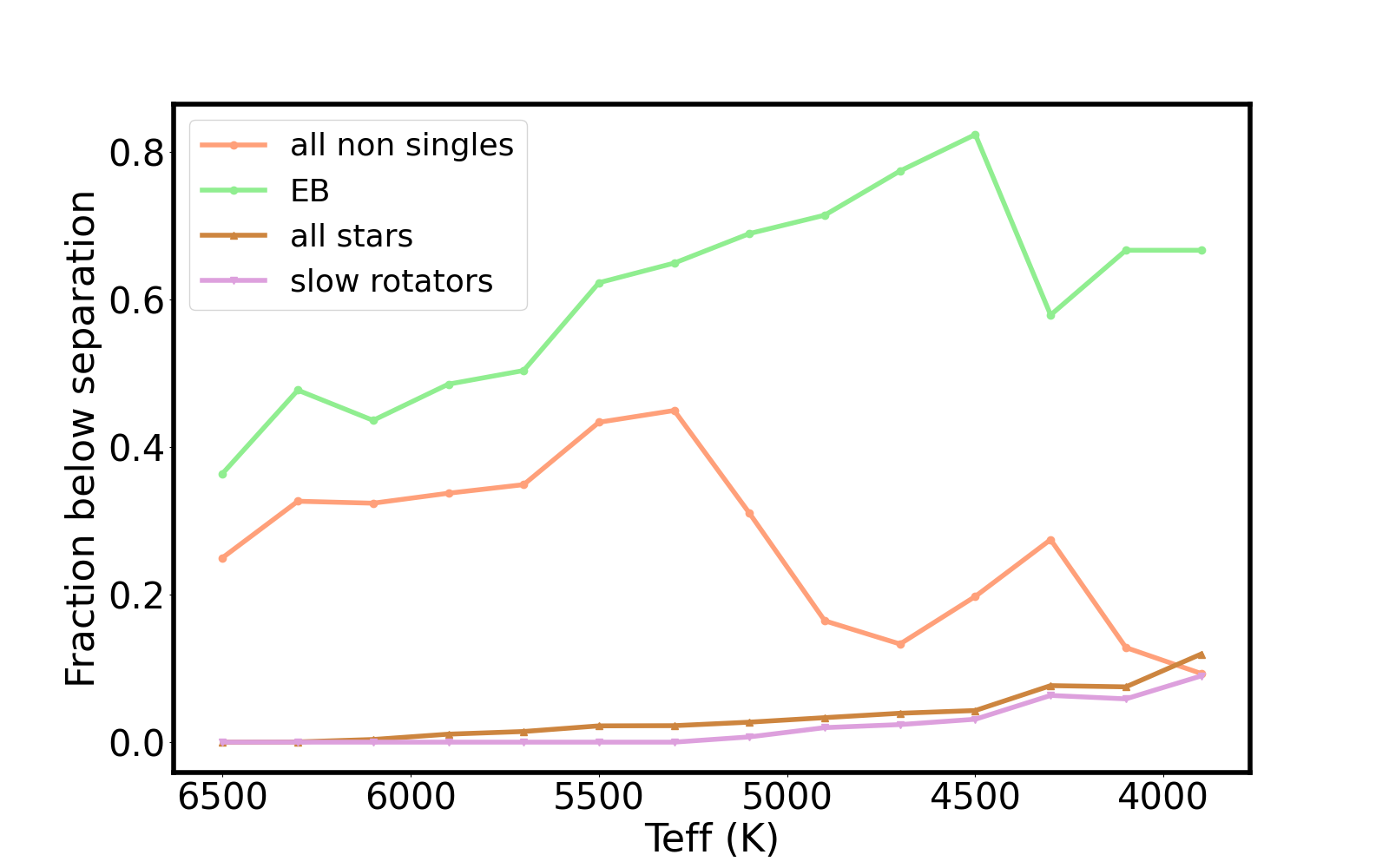}  
    \end{minipage}
    \caption{Each curve shows the fraction of stars below the separation line shown in Figure \ref{fig:cluster_p} for different samples. Every point represents the center of the temperature bin used for the calculations. Colors represent different samples; The left panel shows different types of non-single stars. On the right panel: green - eclipsing binaries from Gaia and Kepler (EB), pink - all known non-single stars (all non singles), brown - the entire sample (all stars), purple - all samples with $P_{rot} > 7$ (slow rotators)} \label{fig:separation_fractions}
\end{figure*}

To find a simple representation, we fitted a $5^{th}$ order polynomial to the median periods of the $300$ Myr cluster between $6600 \geq T_{eff} \geq 3800$. We used temperatures hotter than $3800$ since this is roughly its inflection temperature. The resulting fit is marked in gray on the right panel.\\
This simple line gives a first-order approximation to the initial period of main sequence stars. Because of magnetic braking, at every temperature, stellar rotations of non-interacting stars can only shift to larger periods during their lifetime and should only appear above the curve. Stars with a rotation period below the critical curve are, therefore, either an extremely young star or a synchronized binary. One way to test this separation line is by using a sample of known non-single stars. First, we test the sensitivity of this method for different types of binaries. Figure \ref{fig:separation_fractions} shows in the left panel, the fraction of points below the line for different types of non-single solutions. We see a clear difference between astrometric binaries and eclipsing binaries - eclipsing binaries show a high fraction that decreases with temperature, while other solutions show zero or very low fractions. This difference should not be surprising if we remember the detection biases discussed in section \ref{sec:data}: astrometric solutions require minimal separation to be identified. This translates to relatively wide binaries and, therefore, a low probability of fast and synchronized binaries. We conclude that detections using the separation line are dominated by eclipsing binaries. The right panel of Figure \ref{fig:separation_fractions} shows the fraction of stars below the separation line for both non-singles and general samples. First, we note that there is almost no difference between slow rotators ($P_{rot} > 7$ Days) and general stars, which reflects the fact that the general sample is dominated by slow rotators. Next, we note that although the trend of decreasing fraction with temperature appears in all samples (the jump in the pink curve at $5000$ K is due to the addition of Gaia eclipsing binaries), the difference between non-single stars, and especially eclipsing binaries, to general stars is very significant and emphasizes the efficiency of the method.

While Figure \ref{fig:separation_fractions} supports the validity of the separation line, at least qualitatively, we want to validate it through another independent approach. To do that, we use a magnitude displacement approach, as we describe in the following. 
\subsection{Validation by Magnitude displacement}\label{sec:kmag_diff}
\begin{figure*}
    \centering
    \begin{minipage}[b]{0.45\textwidth}
    \includegraphics[width=\textwidth]{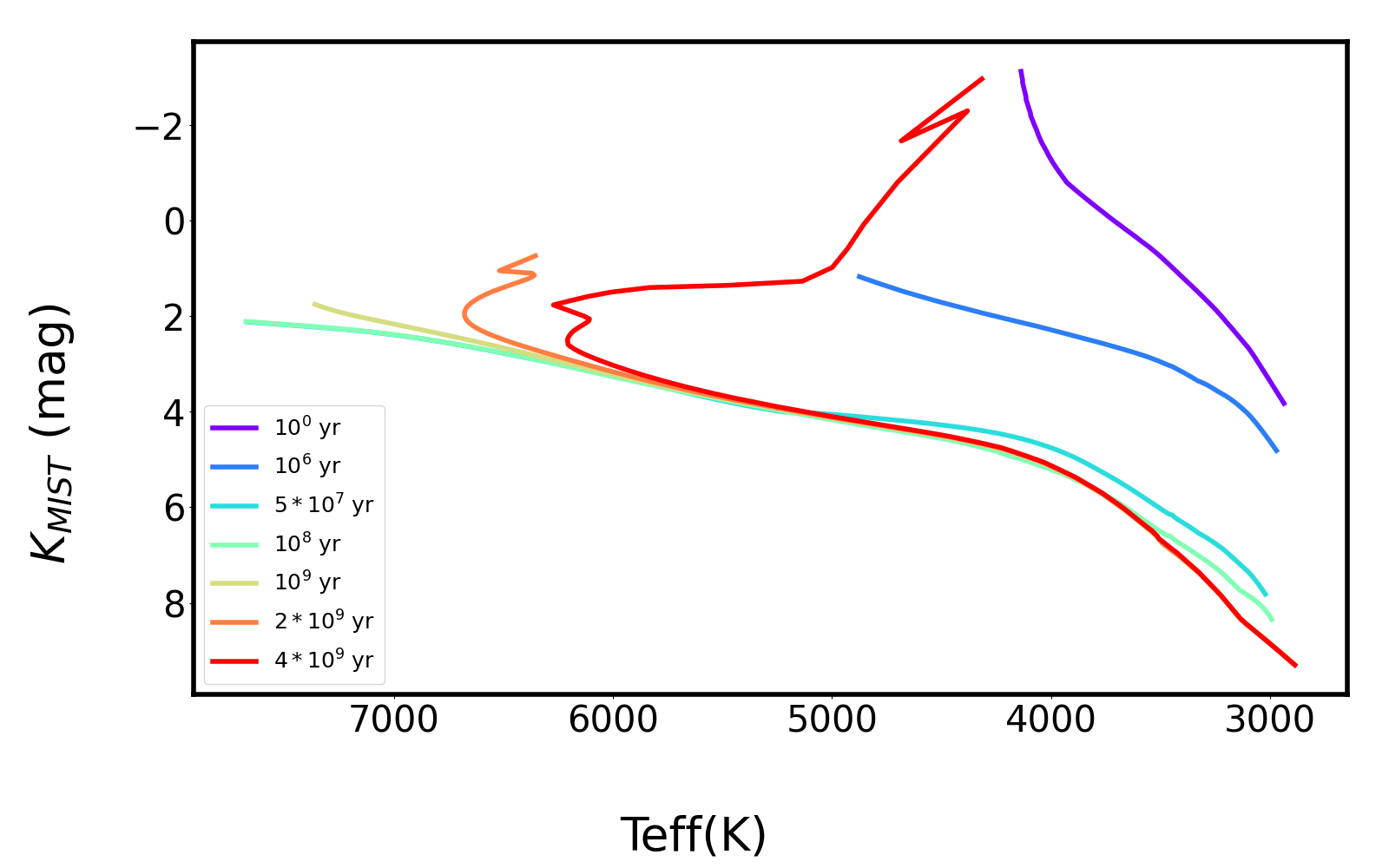}
    \end{minipage}
    \begin{minipage}[b]{0.45\textwidth}
    \includegraphics[width=\textwidth]{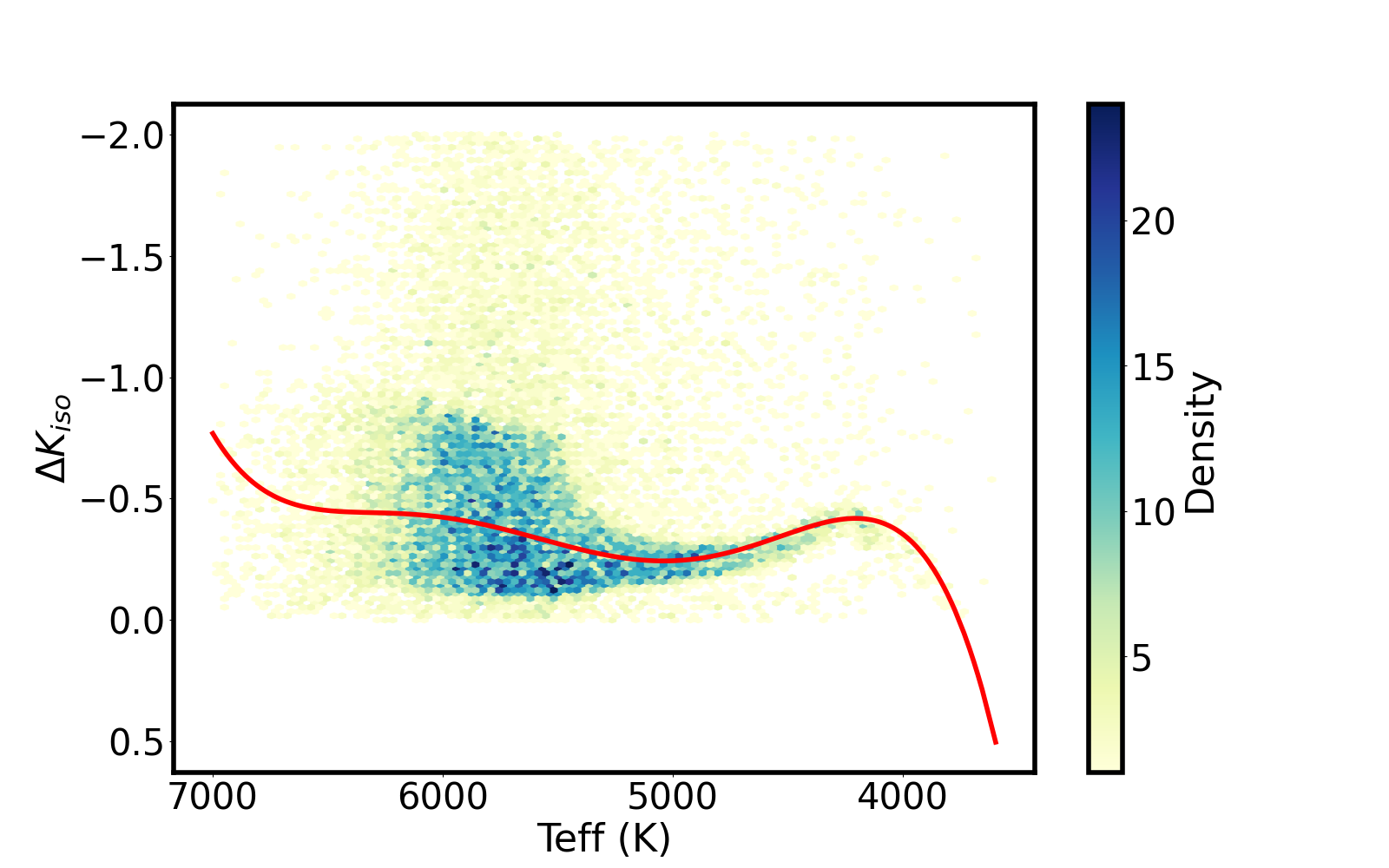}
    \end{minipage}
    \caption{Left -MIST magnitude vs. $T_{eff}$. Right - $\Delta K_{iso}$ vs. $T_{eff}$. The red line represents the maximum density $+ 0.03$ mag, and represents the main sequence line. See text for details.} \label{fig:kmag_density}
\end{figure*}

Generally, a higher-multiplicity system would appear brighter compared to a single star with the same properties, given the luminosity contributed by the companions (in the case of non-compact luminous, typically MS companions) \citep{Haffner1936, Bettis1975, Mermilliod1992}. Therefore, an additional light contribution from a MS companion is expected to increase with an increased mass of the companion. \cite{simonian2019} found binary candidates by subtracting a theoretical isochrone magnitude from the observed K-band magnitude and looking for samples with excess magnitude. We follow a similar procedure for stars with periods from \cite{kamai2024}; we used MIST \citep{Dotter2016, Choi2016} to derive the k-band magnitude of a single star model. MIST evolutionary tracks are characterized by mass and metallicity. Since the measurements of stellar mass might be biased in the case of binarity, instead of using the measured mass from \cite{Berger_2020}, we calculated the main sequence single star mass ($M_{mss}$) from the temperature using mass-luminosity and mass-radius relations of $L \propto M^4$ and $R \propto M^{0.8}$ respectively. We then interpolated the tables based on $M_{mss}$ and $FeH$ to calculate the expected k-band magnitude on a $1$ Gyr isochrone. The left panel of Figure \ref{fig:kmag_density} shows that the specific choice of age is not significant for a wide range of ages and temperatures. It shows MIST results for different ages and solar metallicity. For $3000 < T_{eff} < 6000$ K, only very young stars ($\leq 50$ Myr) shows significantly different magnitudes. For higher temperatures, age effects are more significant as older stars start leaving the main sequence. 

Next, we want to remove evolved stars. To do so, we first calculated the expected lower and upper bounds on $logg$ for different temperature bins using MIST. To remain conservative, we added a width of $0.3$ magnitude on each side and removed all the samples outside the bounds. In addition, we used only samples with solar-like metallicities ($-0.05 < FeH < 0.05$) to reduce the metallicity effects on the main sequence.  We then calculated the difference between the absolute magnitude (using the apparent magnitude from 2MASS and the distance from \cite{Berger_2020}) and the MIST magnitude. We call this quantity $\Delta K_{iso}$. We note that after filtering evolved stars by $logg$, there are still samples with large negative and positive $\Delta K_{iso}$. While this might point to the non-completeness of the filtering process, it might also be a result of errors in measuring the stellar parameters. Since we focus on the identification of non-single stars through magnitude excess, we took only samples with $-2 < \Delta K_{iso} < 0$. This left us with 19264 samples. 

To identify the 'main-sequence' curve in the $T_{eff}-\Delta K_{iso}$ plane, we binned the data into $25$ temperature bins in the range $3800 \leq T_{eff} \leq 7000$. Every temperature bin was binned into $25$ $\Delta K_{iso}$ bins and the bin with maximal sample density was taken as the main sequence $\Delta K_{iso}$. This assumes that the majority of stars are single stars, which aligns with observations of binary fraction \citep{Raghavan2010}; moreover, even for binary stars, systems in which the secondary is of much lower mass would contribute little and appear almost as a single star. We fitted a $5^{th}$ order polynomial to the temperature and maximal density $\Delta K_{iso}$ bins to create a boundary line between single and potential non-single stars. Since other effects can alter the main sequence (rotation, for example), it is better to assign some width to the main sequence curve. In our case, the width takes the form of an extra buffer to the boundary line. We used a buffer of $0.03$ magnitude. The right panel in Figure \ref{fig:kmag_density} shows a density plot of our sample in the $T_{eff}-\Delta K_{iso}$ plane, together with the boundary line.

It is important to note that this separation between single and non-single stars is not expected to be complete. Regardless of the incompleteness of the boundary line itself, we expect to find both populations on both sides of the boundary. For example, a binary with a low-mass companion would show little to no magnitude excess. On the other hand, an old high-mass single star would appear more luminous than expected by a $1$ Gyr isochrone. However, we expect to have statistically more non-single stars above the boundary.

\begin{figure*}
    \centering
    \includegraphics[width=0.6\textwidth]{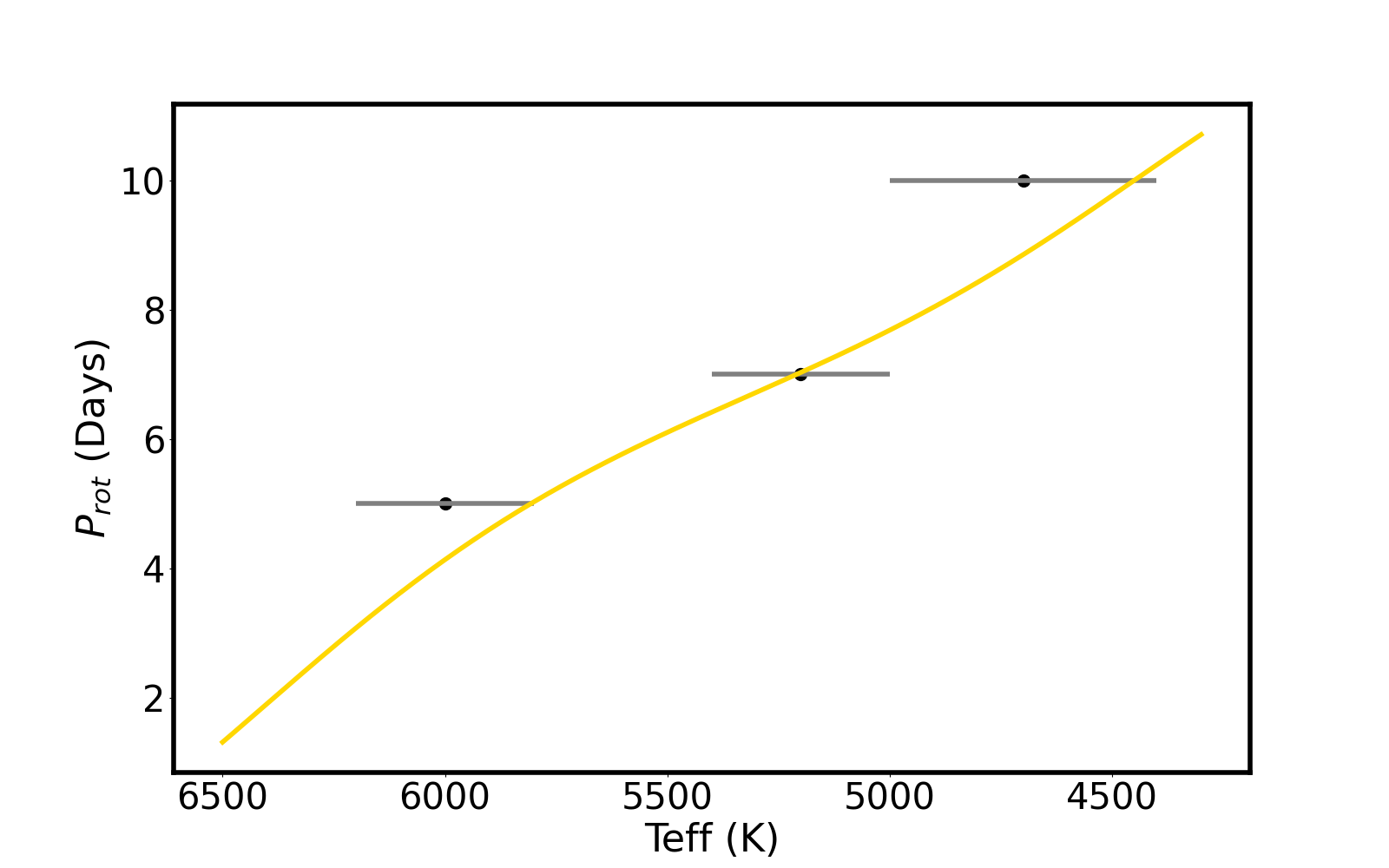}
    \caption{$P_{crit}$ as a function of $T_{eff}$ are marked as circles. Error bars are due to the temperature bin size. The golden curve is the separation line from Figure \ref{fig:cluster_p} } \label{fig:kmag_diff_p_all}
\end{figure*}

Next, we binned the data into temperature bins. For each bin, we count the fraction of samples above the boundary line for all $P_{rot} < P_{cutoff}$ for some $P_{cutoff}$. We used $4 \leq P_{cutoff} \leq 50$ days. Since at large $P_{cutoff}$ the fraction was constant and independent of the period, we normalized the fractions relative to the fraction at $P_{cutoff}=50$ days. To find a value for $P_{cutoff}$ that represents a population dominated by non-single stars, we look for the first value of $P_{cutoff}$ above some threshold. The threshold was found using the same procedure for known non-single stars - here, we expect the fractions to be independent of the period, and the maximum deviation above $1$ was taken as a threshold, which was found to be $1.06$. We, therefore, assign the first period where the fraction of samples above the line is more than $1.06$ (after normalized to the ratio at $P_{cutoff}=50$), as $P_{crit}$. This reflects a temperature-dependent period that potentially marks the transition from a mixture of singles and non-singles ($P_{rot} > P_{crit}$) to a non-singles-dominated population ($P_{rot} < P_{crit}$). Figure \ref{fig:kmag_diff_p_all} shows the found $P_{crit}$ as a function of the center temperature bin, together with the cluster-based separation line from Figure \ref{fig:cluster_p}. Since hot stars ($T_{eff} > 6000$) did not cross the threshold, there are only three points. Nevertheless, we see a good agreement between the points and the separation line, taking into account the width of the temperature bins. We also see agreements with the values found by \cite{simonian2019}. They found a critical value of $7$ days for stars cooler than $5150$ K, which also fits the line. As a last sanity check, we tested all the $217$ potential non-single stars found by \cite{simonian2019} with the provided $P_{rot}$ (found by \citep{McQuillan2014}), and found that all the samples, except one, are below the separation line. \\

\subsection{Validation using peculiar velocity} \label{subsec:kinematic}
\begin{figure*}
    \centering
    \includegraphics[width=0.6\textwidth]{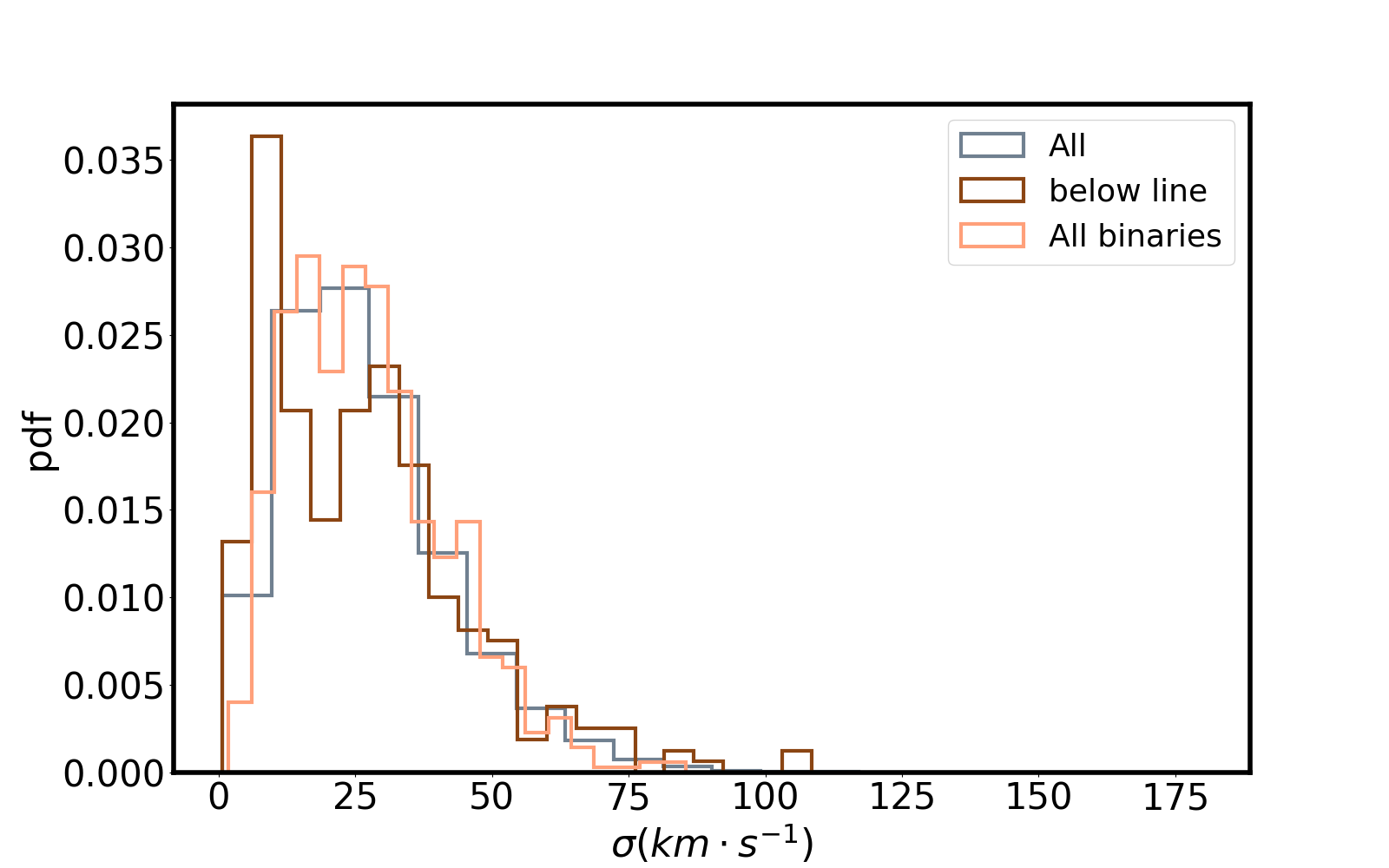}
    \caption{Distributions of $\sigma$ for different data samples. The gray curve represents all the stars with measured $\sigma$, the red curve represents non-single stars. The brown curve represents stars below the separation line. The brown curve shows clear bi-modality around $\sigma \sim 12$ $km \cdot s^{-1}$ } \label{fig:sigma_dist}
\end{figure*}

\begin{figure*}
    \begin{center}
  \begin{minipage}[b]{0.45\textwidth}
    \includegraphics[width=\textwidth]{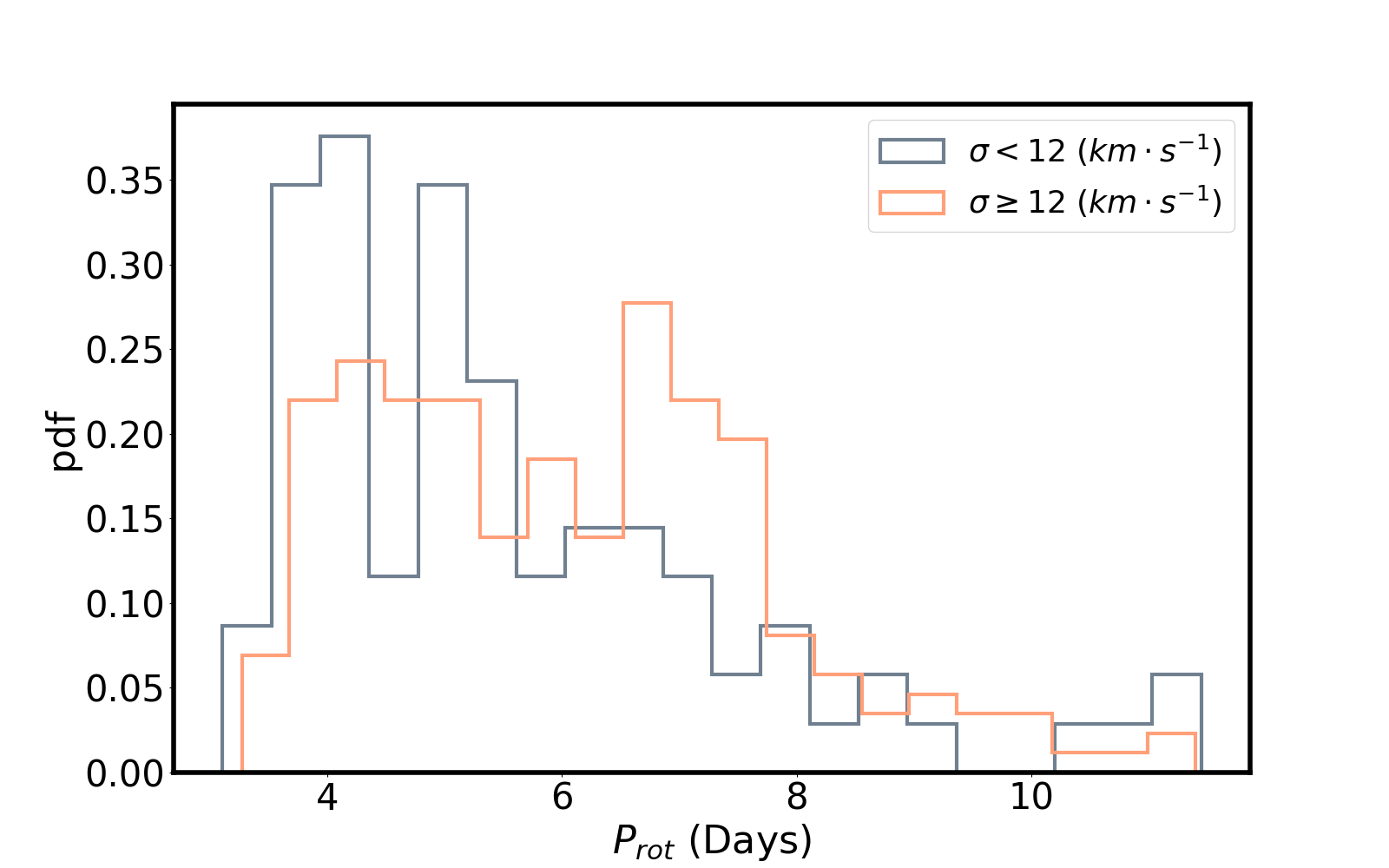}
  \end{minipage}
  \begin{minipage}[b]{0.45\textwidth}
    \includegraphics[width=\textwidth]{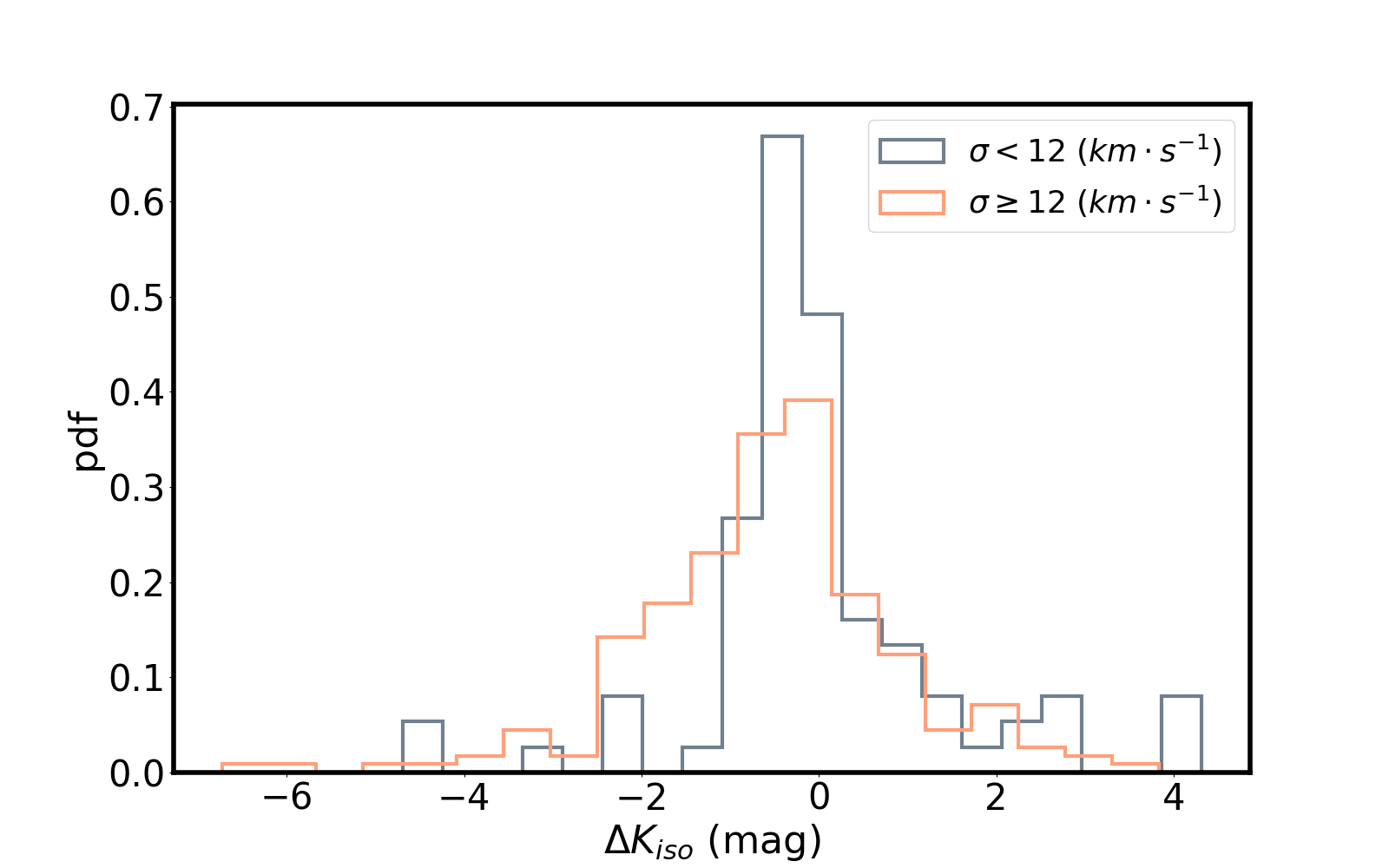}
  \end{minipage}
\caption{distributions of $Prot$ (left) and $\Delta K_{iso}$ for stars below the separation line. The sample is separated to stars with $\sigma < 12 \, km \cdot s^{-1}$ (gray) and stars with  $\sigma \geq 12 \, km \cdot s^{-1}$ (red)} \label{fig:kinematic_prot_kmag_dist}
    \end{center}
\end{figure*}

\begin{figure*}
    \centering
    \includegraphics[width=0.6\textwidth]{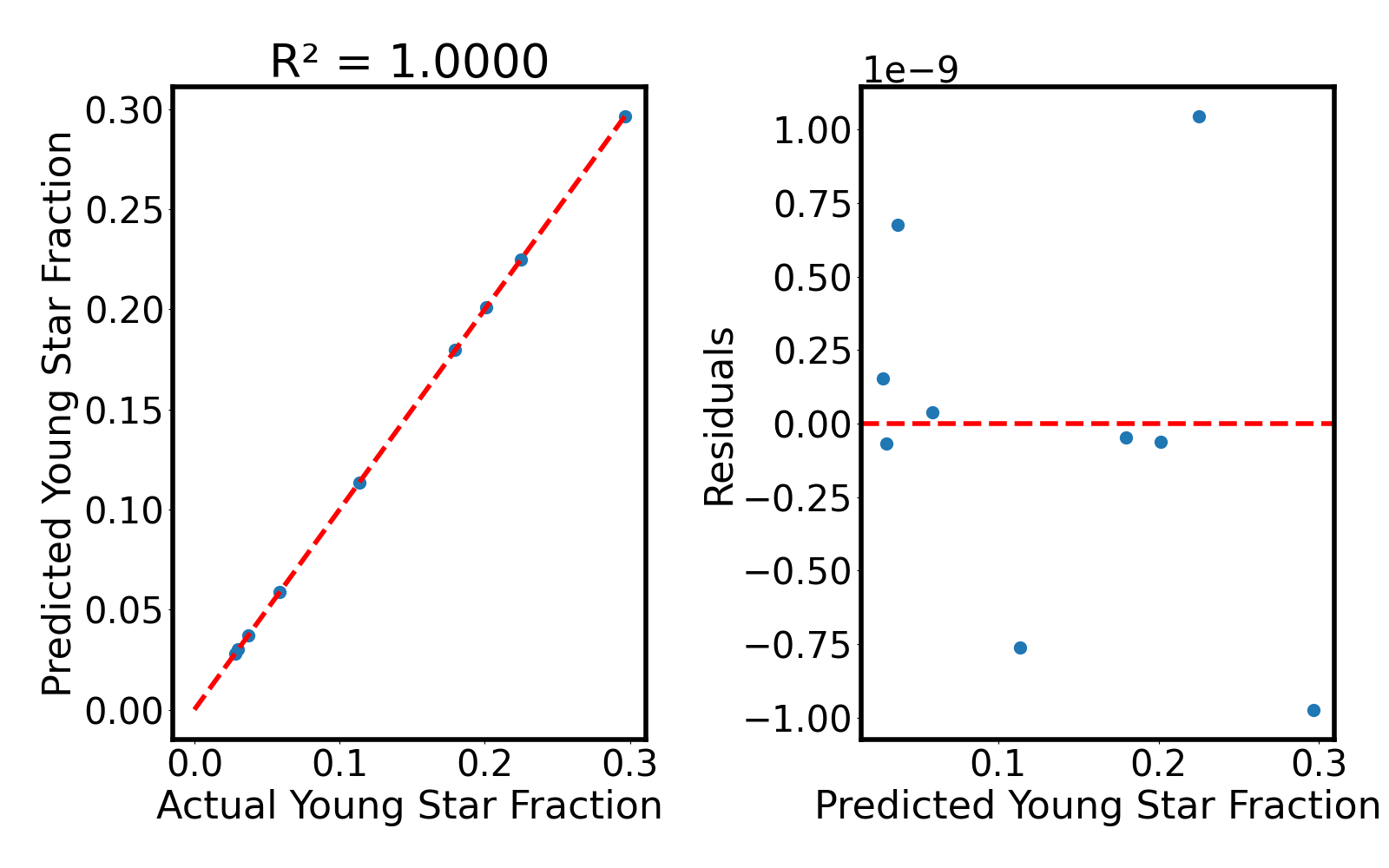}
    \caption{Left - True young star fraction vs predicted young star fraction according to equation \ref{eq:prob_teff_p_fit}. The red line shows the $x=y$ line. Right - residuals of the plot on the left panel. } \label{fig:young_probs_teff_prot}
\end{figure*}

\begin{figure*}
    \centering
    \begin{minipage}{0.45\textwidth}
    \includegraphics[width=\textwidth]{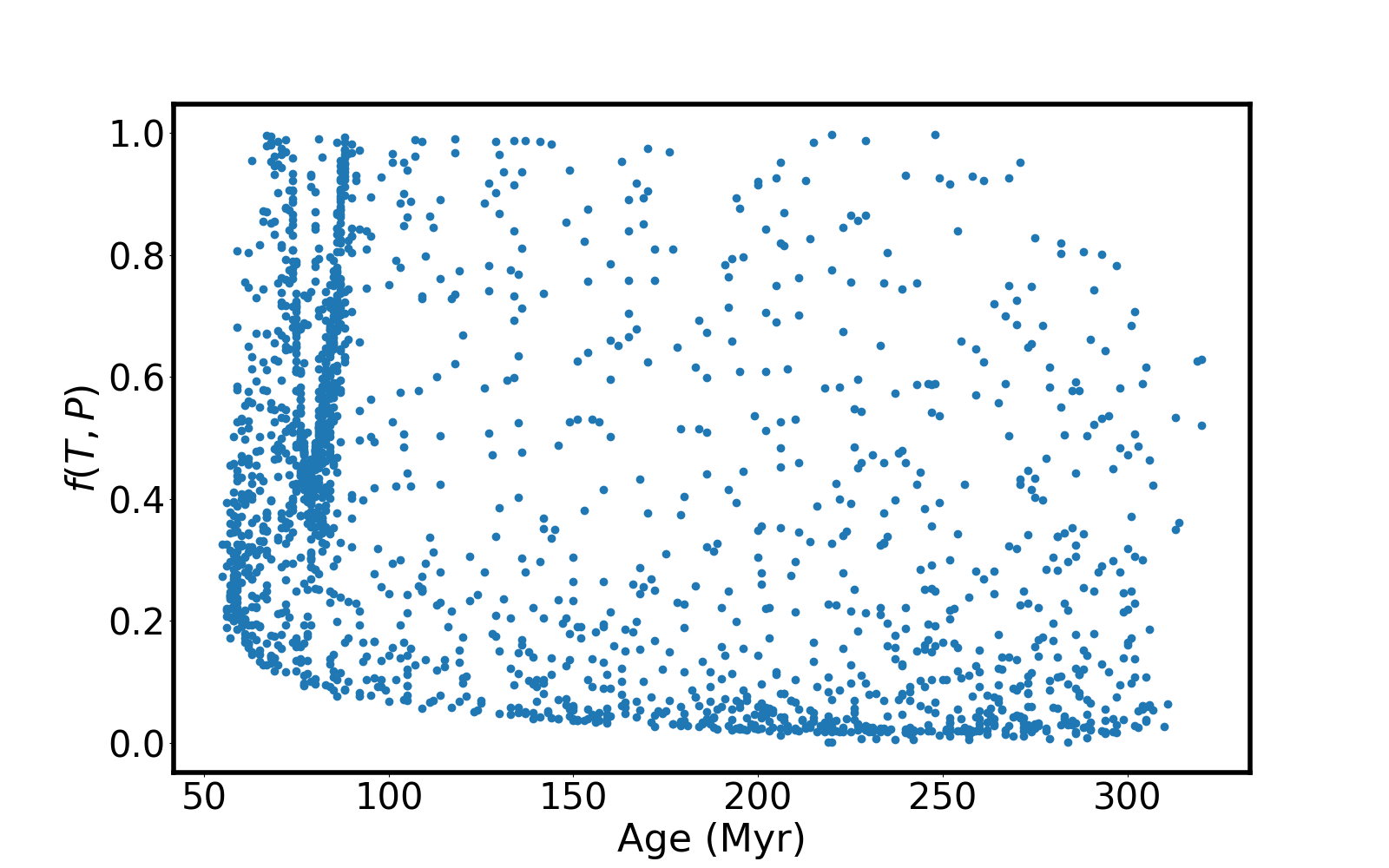}     
    \end{minipage}
    \begin{minipage}{0.45\textwidth}
    \includegraphics[width=\textwidth]{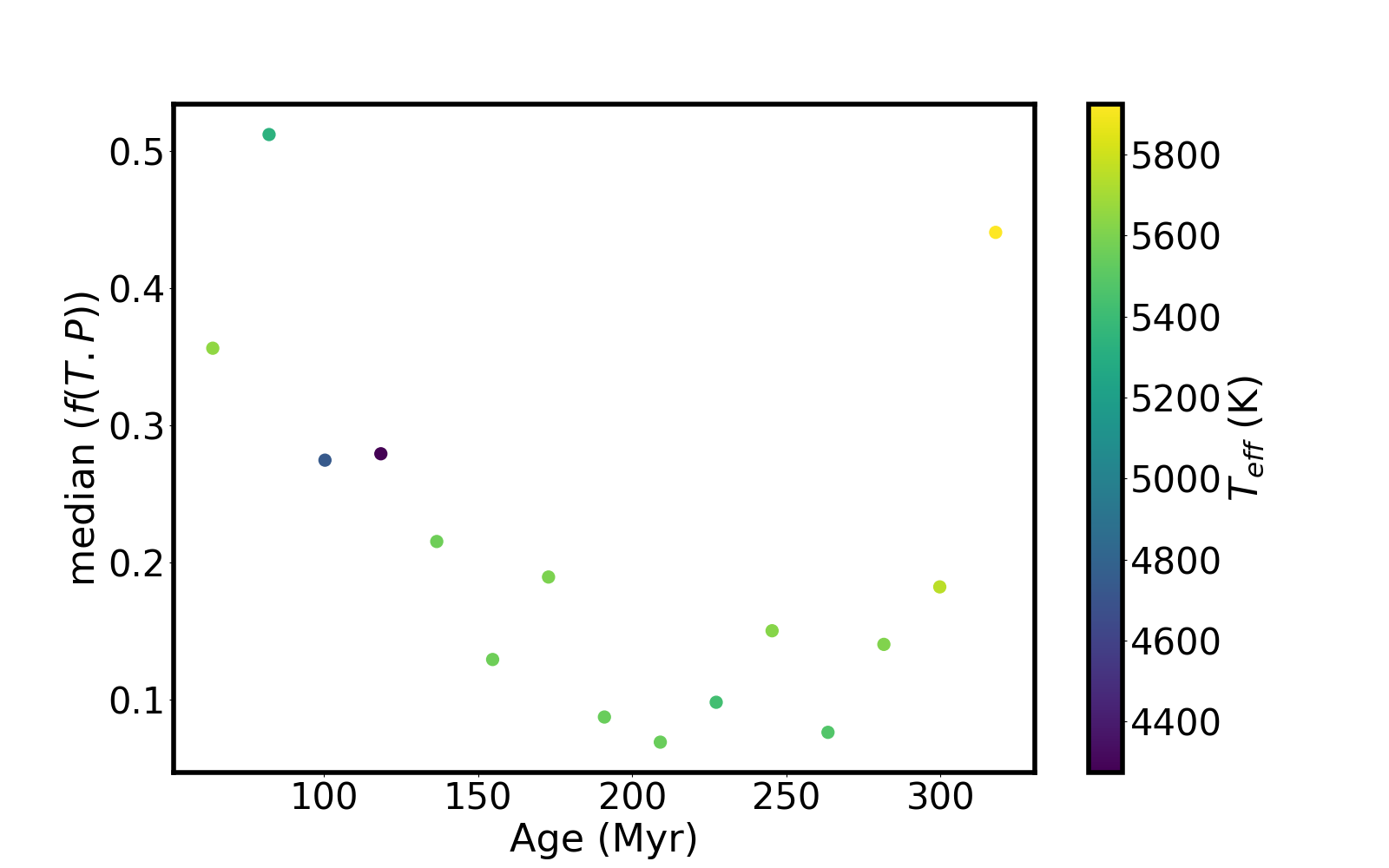}     
    \end{minipage}
    \caption{comparison of gyrochronology ages from \cite{Bouma2024} vs. young star fraction prediction ($f(T,P)$. The left panel shows a scatter of all points and the right panel shows the median $f(T,P)$ over age bins. Only samples below the separation line are shown. } \label{fig:bouma_age_prob}
\end{figure*}

Although we showed a correlation between our method and the magnitude displacement method, there is still one caveat with the proposed method - stars below the separation line might be very young, and without the ability to separate very young stars from general stars, it is hard to asses the efficiency of the method. Infering the age from the period, as in gyrochronology, is problematic because it is biased by the binary population (see, for example \cite{Daher2022}). On the other hand, using isochrone fitting for young ages ($< 1$ Gyr) is highly inaccurate, as this method is precise at temperatures and luminosities that separate stars from the main sequence and usually have large error bars. For this reason, we analyze the peculiar velocities as a proxy for kinematic age. It is known that the kinematic properties of stars change during their lifetime \citep{stromberg1946}, which makes them useful as age indicators \citep{nordstrom2004, Holmberg2007, Holman1999, Yu2018,  Lu2021, Chen_2021}. Kinematic age is usually valid only in a statistical manner and not per-star, but in our case, it is not a problem. If a sample of stars that lie below the separation line consists of a combination of very young stars and general synchronized binaries, the two populations would have different kinematic properties. To test this, we use the catalog of kinematic properties provided by \cite{Chen_2021}, who provided kinematic properties to $35835$ Kepler stars. Specifically, we look at the root mean square of the peculiar velocities - $\sigma = \sqrt{\frac{1}{3}(v_x^2+v_y^2+v_z^2)}$. When cross-matched with our sample, we were left with  $15572$ samples, out of which $296$ are below the line. We also cross-matched the kinematic catalog with known non-single stars, resulting in $836$ stars. Figure \ref{fig:sigma_dist} shows the $\sigma$ distributions of the entire sample (gray), all non-single stars (red), and stars below the separation line (brown). The distributions of the entire sample and general binaries show very good agreement. Their Kolmogorov-Smirnov (KS) test p-value is $0.25$. However, the distribution of samples below the line looks very different with a clear bi-modality around $\sigma \sim 12 km \cdot s^{-1}$. The p-value of the KS test between the brown and gray samples is $1.66 \cdot 10^{-5}$. This clear difference suggests that we indeed observe two distinct populations - very young stars (with $\sigma < 12 \, km \cdot s^{-1} $) and synchronized binaries that follow the general $\sigma$ distribution. Another verification of the bi-modality of the sample is shown in Figure \ref{fig:kinematic_prot_kmag_dist}. It shows the distributions of $P_{rot}$ and $\Delta K_{iso}$ for the sample of stars below the separation line. The sample is further divided into stars with $\sigma < 12 \, km \cdot s^{-1}$ (gray), and stars with $\sigma \geq 12 \, km \cdot s^{-1}$ (red) and we see very different distributions in both parameters. Moreover, the sample with $\sigma \geq 12 \, km \cdot s^{-1}$ is more luminous (lower $\Delta K_{iso}$) as one expects from binaries-dominated sample. We can conclude that stars below the separation line with $\sigma > 12 \, km \cdot s^{-1} $ are binaries with very high confidence. Therefore, knowing the peculiar velocity of stars together with temperature and period gives a very accurate binarity prediction. However, peculiar velocity is not always available. One can estimate a $\sigma$-free probability that can be applied to every star below the line. We will base such probability on the deviation of the brown curve from the general distribution. For that, we measure the 'excess' of the brown distribution at $\sigma < 12 \, km \cdot s^{-1}$ compared to the general distribution, i.e. effectively measuring the difference in the areas of the two distribution curves. This is done by calculating the expected number of stars with $\sigma < 12 \, km \cdot s^{-1}$ for a general sample (with the same total number as the brown curve), subtraction of the true number of stars from the expected number, and normalizing by the total number of stars in the brown curve. We perform the above calculation on 2-dimensional bins of temperature and period and fit a polynomial function of two variables. To reduce the probability of overfitting, we limit the maximal polynomial degree to 3, take only bins with at least $10$ points in the sample below the line, and take only bins with excess distribution. This results in $9$ data points. The best-fit was found to be the following third-order polynomial function:
\begin{equation}\label{eq:prob_teff_p_fit}
\begin{split}
    f(T,P) = & -8 \cdot 10^{-6} - 0.0926 T - 0.0533P \\
             & - 0.0012TP - 1.4 \cdot 10^{-5} T^2 \\
             & -0.9806P^2 + 1.54 \cdot 10^{-4}TP^2 \\
             & + 0.0133P^3,
\end{split}
\end{equation}
Where $T$ is the temperature in Kelvin and $P$ is the period in days.
The left panel of Figure \ref{fig:young_probs_teff_prot} shows the predictions of young star fractions, using Eq. \ref{eq:prob_teff_p_fit}, vs. the true fractions, for the 9 points used for fitting. The right panel shows the residuals. It can be seen that the predictions show excellent agreement with the true values. To validate $f(T,P)$, we use gyrochronological ages calculated by \cite{Bouma2024}. Since gyrochronology is less accurate on young stars (because of the large binarity fraction), we expect a large scatter. However, we can expect a statistically negative correlation between age and $f(T,P)$. Figure \ref{fig:bouma_age_prob} shows a scatter of all samples (left panel) and the median $f(T,P)$ over age bins (right panel). The Figure matches our expectation for a large scatter and a statistically significant correlation between older stars and lower $f(T,P)$. Interestingly, we see an increase in $f(T,P)$ around $300$ Myr, which is exactly the boundary used to construct the separation line.
To use Eq. \ref{eq:prob_teff_p_fit} to assign a probability for non-single candidates, we need to remember that $f(T,P)$ is not a proper probability function. specifically, it is not bound and can get arbitrarily higher and lower values. To convert it to a proper probability function, we normalize it to be in the 0-1 range:
\begin{equation} \label{eq:min_max_norm_prob}
    f_{norm}(T,P) = \frac{f(T,P) - f_{min}(T,P)}  {f_{max}(T,P) -  f_{min}(T,P)},
\end{equation}
Where $f_{min}(T,P)$ and $f_{max}(T,P)$ are the minimum and maximum values of $f$ over the entire samples. Using Eq. \ref{eq:min_max_norm_prob}, we can calculate the $\sigma$-free probability of a star to be a non-single star, given that this star is below the line as:
\begin{equation} \label{eq:b_prob}
    P(nss | T,P) = 1 - f_{norm}(T,P)
\end{equation}

\section{Results} \label{sec:results}
\subsection{Identifying non-single systems} \label{subsec:non_singles_catalog}
We now use the separation line to find potential non-single stars in the \textit{Kepler} field. We use the catalog from \cite{kamai2024}, and choose all the stars with $P_{rot}$ and $T_{eff}$ such that they sit below the critical separation line. For each sample, we specify $\sigma$ when available, and assign the $\sigma$-free probability of being a non-single according to Eq. \ref{eq:b_prob}. We emphasize the fact that this probability is a result of statistical analysis of the over-density of stars with low peculiar velocity, and since there is some overlap between the distributions of non-singles and young stars, it does not reflect the real, unknown, probability of non-single stars below the line. Nevertheless, it measures samples that are more likely to be non-singles than others. In this light, we want to remove stars that are known to be young, regardless of their $\sigma$ or $P(nss|T,P)$. First, we removed stars that appear in young stellar clusters. Second, we used the ages calculated by \cite{Bouma2024} for planets. They used gyrochronology and lithium abundance as two independent age estimators. While gyro alone may not be accurate, Lithium is not sensitive to binarity, and when both methods predict an age lower than $300$ Myr, the confidence of the prediction is higher. To remain conservative, we removed those samples. This left us with $2229$ samples. The final sample should be understood as follows: if $\sigma$ is available and we have $\sigma > 12$, we can classify the sample as a non-single star. If $\sigma \leq 12$, it is probably a young star but still might be non-single. The evolution of young single stars and young binaries might be different \citep{Meibom2007}, but we expect a non-trivial mixture of non-singles and singles in this regime. If $\sigma$ is not available, we can use $P(nss|T,P)$ as a general probability for the sample to be non-single. An example of the resulting catalog is shown in Table \ref{table:binaries_table}. The full table is available in machine-readable format.

\begin{table*}
\centering
\begin{tabular}{||c|c|c|c|c|c|c|c|c|c||}
    \hline
    \textbf{KID} & \boldmath$P_{rot}$ (Days) & \textbf{Teff} & \textbf{logg} & \textbf{FeH} & \boldmath$\Delta K_{iso}$ & \boldmath$k_{mag}$ & \boldmath$K_{MIST}$ & $\sigma \, (km \cdot s^{-1})$
        & \boldmath$P(nss | T,P)$ \\ \hline
   6956701 & 4.102 & 3896.1 & 1.48 & 0.244 & -6.451 & -3.759 & 2.692 & - & 0.297 \\ \hline
        10470116 & 7.545 & 5035.6 & 4.518 & 0.041 & -2.361 & 1.871 & 4.231 & - & 0.748 \\ \hline
        3733157 & 4.19 & 5514.3 & 4.287 & 0.017 & -0.683 & 3.086 & 3.77 & - & 0.741 \\ \hline
        3752144 & 4.693 & 5174.9 & 4.442 & 0.087 & -0.446 & 3.671 & 4.117 & - & 0.729 \\ \hline
        10604021 & 7.804 & 4955.6 & 2.889 & -0.438 & -1.579 & 1.649 & 3.227 & - & 0.751 \\ \hline
        7049035 & 4.623 & 5210.3 & 4.311 & 0.235 & -0.857 & 3.278 & 4.135 & - & 0.731 \\ \hline
        5536695 & 4.166 & 5796.4 & 4.43 & -0.033 & -0.152 & 3.319 & 3.471 & - & 0.736 \\ \hline
        2436332 & 4.534 & 4342.1 & 2.008 & -0.004 & -4.964 & -2.754 & 2.21 & - & 0.537 \\ \hline
        7294914 & 5.942 & 5430.5 & 3.788 & -0.016 & -2.21 & 1.629 & 3.839 & - & 0.745 \\ \hline
        2284919 & 10.124 & 3762.7 & 4.705 & 0.189 & -0.394 & 5.482 & 5.876 & - & 0.793 \\ \hline
        6129696 & 4.137 & 5799.7 & 3.901 & 0.012 & -1.71 & 1.776 & 3.486 & - & 0.736 \\ \hline
        6465729 & 7.536 & 4387.1 & 4.626 & -0.003 & -0.342 & 4.584 & 4.926 & - & 0.693 \\ \hline
        8195444 & 4.258 & 4961.5 & 3.43 & -0.058 & -0.231 & 2.473 & 2.703 & - & 0.697 \\ \hline
        1849430 & 7.759 & 4811.2 & 4.553 & 0.062 & -0.325 & 4.13 & 4.454 & - & 0.748 \\ \hline
    \hline
\end{tabular}
\caption{potential non-single stars. $\sigma$ refers to the peculiar velocity, when available. $P_{nss}(T,P)$ is the probability of being non-single according to equation \ref{eq:b_prob}. The full table is available in machine-readable format.}
\label{table:binaries_table}
\end{table*}

\subsection{Identifying triple star systems} \label{subsec:triples}
\begin{figure*}
    \centering
    \includegraphics[width=0.55\textwidth]{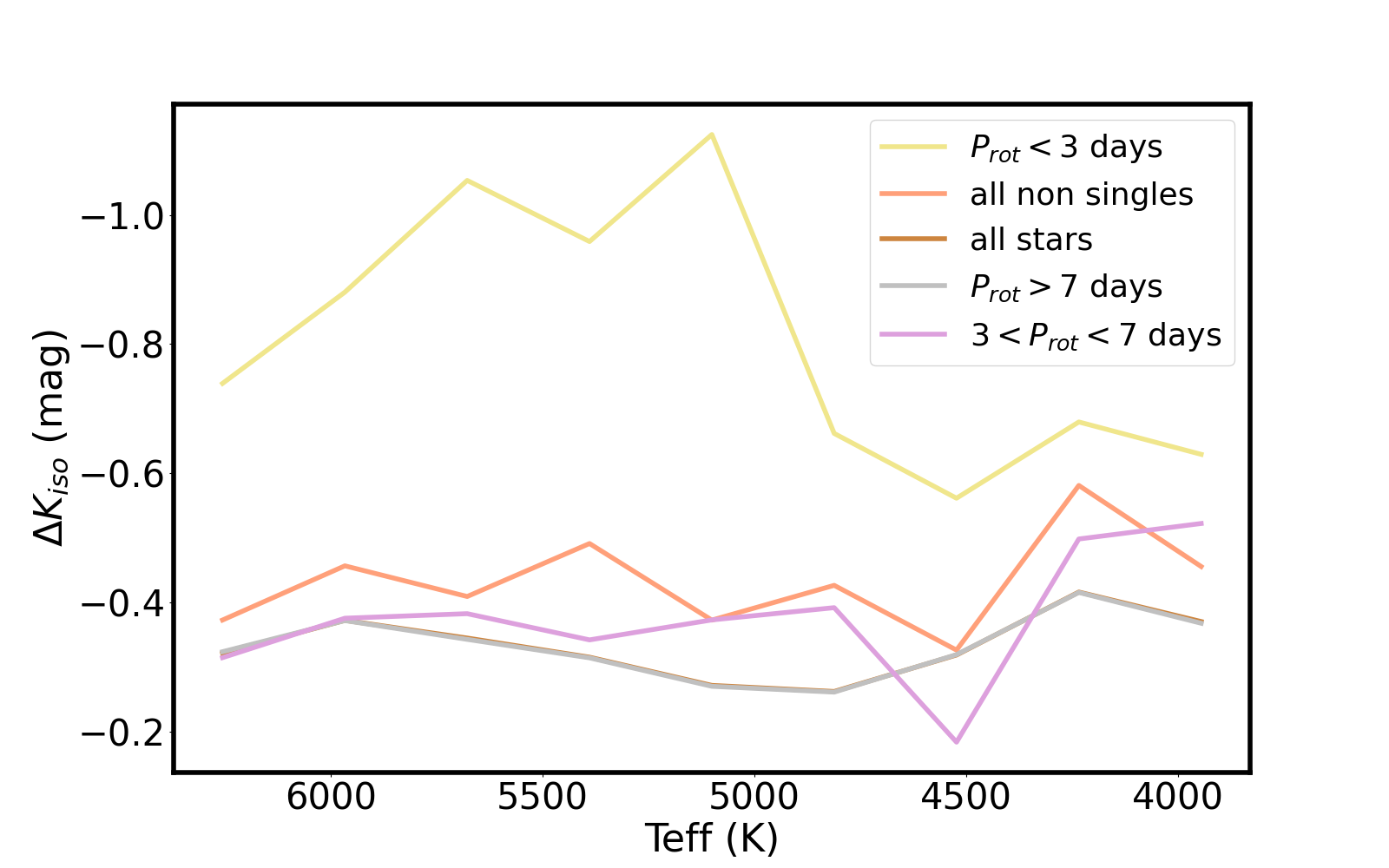}
    \caption{median $\Delta K_{iso}$ as a function of temperature for different samples. Every point represents the center of the temperature bin used for the calculations. Colors represent different samples; yellow -  $P_{orb} < 3$, pink - all known non-single stars (all non singles), brown - the entire sample (all stars), gray - all samples with $P_{rot} > 7$, purple - all samples with $3 < P_{rot} > 7$. The brown curve aligns with the gray curve and is therefore hardly seen.} \label{fig:kmag_triples}
\end{figure*}

Another interesting implication of period-multiplicity relations is related to triple systems. \cite{Tokovinin2006} analyzed $165$ solar-type spectroscopic binaries and found that the probability of having a third companion is strongly dependent on the orbital period, reaching $~96\%$ for binaries with $P_{orb} < 3$ days. Since for such close separation, the binaries are synchronized, {\emph all stars with $P_{rot} < 3$ days are effectively triple systems}. Since $3$ days is the lower bound for periods in the catalog of \cite{kamai2024}, we took fast rotators from \cite{McQuillan2014}, \cite{Santos2019}, \cite{Santos2021}, and \cite{Reinhold2023}; we took stars with $P_{rot} < 3$ and $3800 < T_{eff} < 6200$ from all those catalogs. The reason for the upper temperature bound is that this is the temperature at which the separation line reaches $3$ days. To validate the distinction of this sample from general stars and general non-single stars, we again look at magnitude displacement. Since the fraction of triples decreases sharply with the orbital period, the sample of non-single stars is dominated by binaries. Moreover, since triples are generally more luminous than binaries, we expect them to have different $\Delta K_{iso}$. Figure \ref{fig:kmag_triples} shows the median $\Delta K_{iso}$ as a function of $9$ temperature bins for different samples. Comparing the potential triples ($P_{rot} < 3$ days) and all non-single stars, we see the expected behavior- the triples show significantly higher $\Delta K_{iso}$. We also see that the $\Delta K_{iso}$ of general stars is lower, again as expected. Looking at general stars, we see no difference between slow rotators ($P_{rot} > 7$) and all stars as both curves are the same. This is because the number of stars with $3 < P_{rot} < 7$ is much lower than the total number of stars and has no statistical effect on the general sample. However, when we explicitly look at stars with $3 < P_{rot} < 7$, we see that their $\Delta K_{iso}$ is almost always between the values of slow rotators and non-single stars, as expected from a 'transition' regime.

\begin{figure*}
    \centering
    \includegraphics[width=0.65\textwidth]{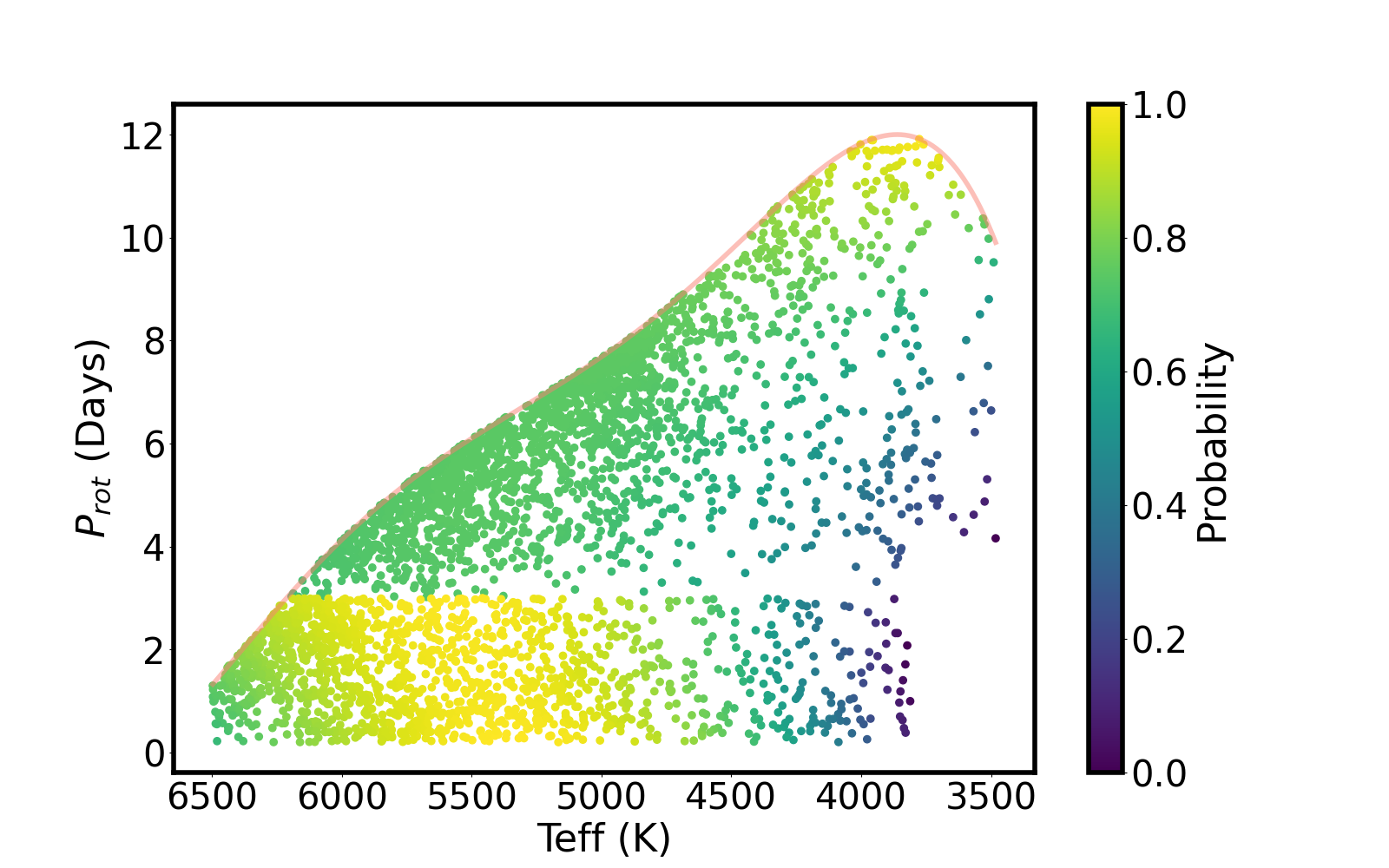}
    \caption{$T_{eff}$ vs. $P_{rot}$ for potential non singles and triples from Tables \ref{table:binaries_table} and \ref{table:triples_table}. Colored points are non-single probabilities calculated using Eq. \ref{eq:b_prob}. The jump in probabilities between the samples is due to different normalizations.} \label{fig:final_catalogs_scatter}
\end{figure*}

To create a catalog of triples, we take all samples with $P_{rot} < 3$ days and $\Delta K_{iso} < 0$. Similarly to the general sample, we removed stars that appear in young clusters, or have young age predictions from both gyro and lithium in \cite{Bouma2024}. This results in $1760$ stars, out of which $1518$ are below the separation line and form our final catalog of potential triple systems. Table \ref{table:triples_table} shows an example of the final catalog.
Figure \ref{fig:final_catalogs_scatter} shows the $P_{rot}$ vs. $T_{eff}$ for both final catalog of potential non-singles from \ref{subsec:non_singles_catalog} and the final triples catalog. The colors of the points refers to $P(nss|T,P)$. We can see the separation between the two datasets at $P_{rot}=3$ Days. This is because the sample used for general non-singles \citep{kamai2024} has a lower bound of three days, while the sample used for triples identification \citep{McQuillan2014, Santos2019, Santos2021, Reinhold2023} has an upper bound of three days. Since we treat the two samples as distinct datasets, we used different normalizations in Eq. \ref{eq:min_max_norm_prob} when calculating their probabilities. This creates the jump in probability at $P_{rot}$ = 3 Days. The average probability of potential non-singles is $0.72$ and the average probability of potential triples is $0.87$.

\begin{table*}
    \centering
    \begin{tabular}{||c|c|c|c|c|c|c|c|c|c|c||}
    \hline
        \textbf{KID} & \boldmath$P_{rot}$ & \textbf{Teff} & \textbf{logg} & \textbf{FeH} & \boldmath$\Delta K_{iso}$ & \boldmath$k_{mag}$ & \boldmath$K_{MIST}$ & \textbf{Reference} &
        $\sigma \, (km \cdot s^{-1})$ &
        \boldmath$P(nss | T,P)$ \\ \hline
        1026474 & 1.569 & 4177 & 4.599 & 0.197 & -0.626 & 4.693 & 5.318 & McQ14 & - & 0.415 \\ \hline
        1572802 & 0.374 & 4035.4 & 4.63 & 0.142 & -0.544 & 4.947 & 5.492 & McQ14 & - & 0.345 \\ \hline
        1872885 & 1.872 & 3936.7 & 4.606 & 0.404 & -0.831 & 4.85 & 5.681 & McQ14 & - & 0.146 \\ \hline
        2300039 & 1.712 & 3829.7 & 4.678 & 0.193 & -0.465 & 5.325 & 5.789 & McQ14 & - & 0.014 \\ \hline
        2436635 & 1.175 & 4733.7 & 4.496 & 0.357 & -0.679 & 3.962 & 4.642 & McQ14 & - & 0.827 \\ \hline
        2442866 & 2.934 & 4921.5 & 4.441 & 0.355 & -0.696 & 3.745 & 4.44 & McQ14 & - & 0.888 \\ \hline
        2557669 & 1.864 & 4081.1 & 4.589 & 0.326 & -0.764 & 4.717 & 5.481 & McQ14 & - & 0.31 \\ \hline
        2985366 & 0.244 & 4260.8 & 4.639 & -0.037 & -0.386 & 4.702 & 5.088 & McQ14 & - & 0.577 \\ \hline
        3130391 & 1.228 & 4233.1 & 4.584 & 0.27 & -0.992 & 4.276 & 5.268 & McQ14 & - & 0.481 \\ \hline
        3430287 & 0.469 & 4320 & 4.645 & -0.06 & -0.322 & 4.671 & 4.993 & McQ14 & - & 0.606 \\ \hline
        3530387 & 0.54 & 4339.3 & 4.658 & -0.136 & -0.182 & 4.748 & 4.93 & McQ14 & - & 0.616 \\ \hline
        3539331 & 1.541 & 4461.3 & 4.604 & 0.008 & -0.359 & 4.469 & 4.828 & McQ14 & - & 0.653 \\ \hline
        3541346 & 0.911 & 4303.6 & 4.674 & -0.189 & -0.096 & 4.858 & 4.953 & McQ14 & - & 0.56 \\ \hline
        3556533 & 1.255 & 4890.4 & 4.432 & 0.416 & -0.747 & 3.74 & 4.487 & McQ14 & - & 0.889 \\ \hline
        \hline
    \end{tabular}
    \caption{Example table of potential triple systems. The full table is available in a machine-readable format. } \label{table:triples_table}
\end{table*}

\subsection{Fast spinning planet Host Stars} \label{subsec:planets}
\begin{figure*}
    \centering
    \includegraphics[width=0.6\textwidth]{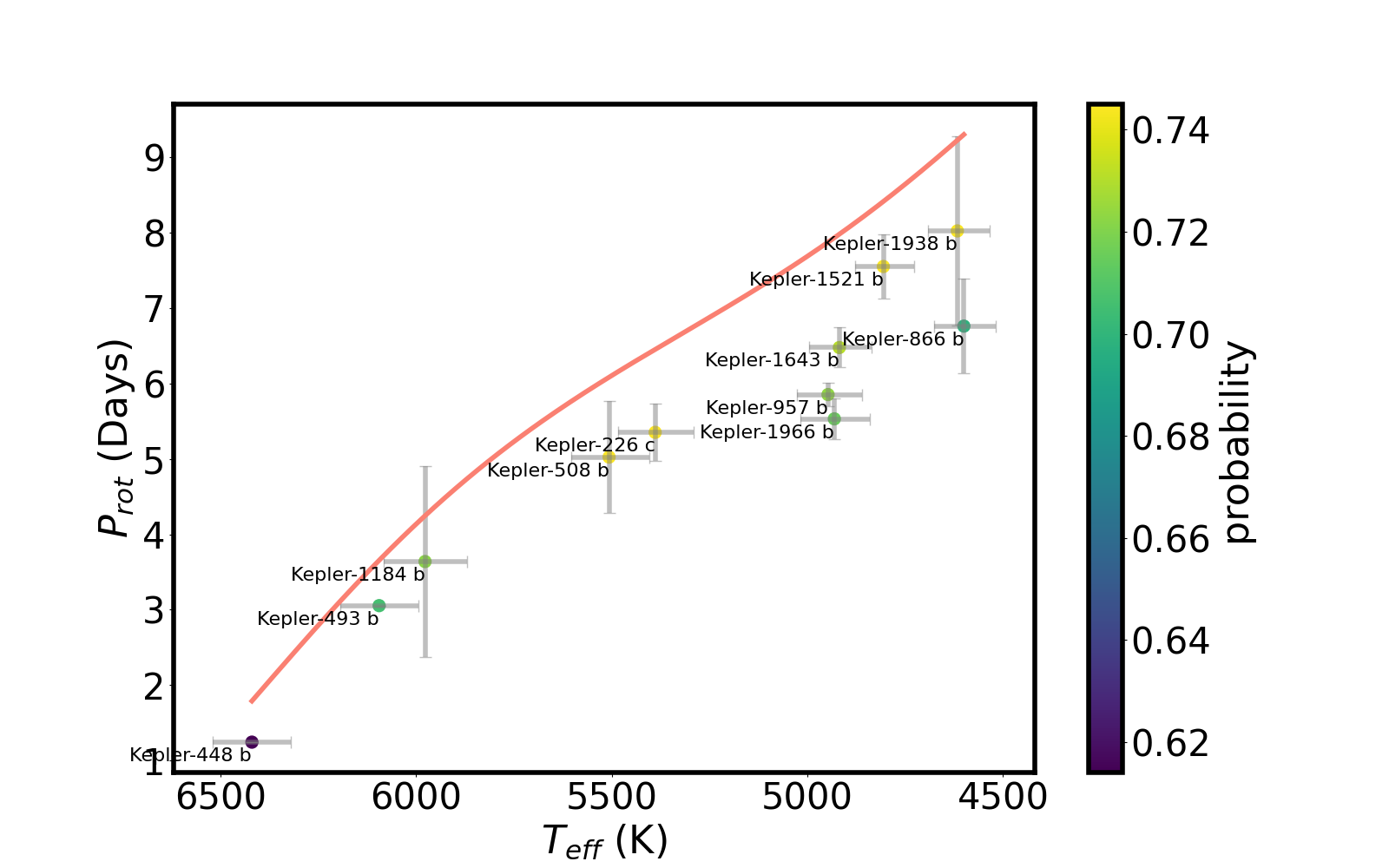}
    \caption{stellar rotation period as a function of $T_{eff}$ for planet host stars. The golden curve is the non-singles separation line derived in section \ref{sec:phot_binaries}.} \label{fig:planet_scatter}
\end{figure*}

It is interesting to apply our findings to planet-host stars. We can use our period-temperature separation line, found in section \ref{sec:phot_binaries}, and see if some planet-host stars are found below the line, possibly making them synchronized binary stars. If so, and if the planet detection is correct, the system is potentially a circumbinary planet system. Those systems are important for understanding both planet and binary formation, and only a small number of such systems have been found to date. As mentioned before, another option is that the system is very young. Similarly to \ref{subsec:non_singles_catalog}, we use the age predictions from \cite{Bouma2024}, when both gyro and lithium ages are available, together with $P(nss|T, P)$, to take into account young systems. \\ 
We cross-matched the catalog from \cite{kamai2024} with the catalog of confirmed planet host stars from \textit{Kepler}. We added stars from \cite{McQuillan_2013}, which calculated periods of planet host stars. We then took only samples below the separation line with $T_{eff} > 3800$ K which results in $11$ systems. Figure \ref{fig:planet_scatter} shows the resulting sample and the separation line. The colors are $P(nss|T,P)$, calculated using Eq \ref{eq:b_prob}. For normalization in Eq. \ref{eq:min_max_norm_prob}, we used $f_{min}$ and $f_{max}$ from the general sample (\ref{subsec:non_singles_catalog}). We see that most samples have a probability close to the average probability of the general sample ($0.73$), with the exception of \textit{Kepler-448} that shows significantly lower probability. The errors in the period were taken from \cite{kamai2024} (we took the total error, which combines consistency error and the model's confidence) and \cite{McQuillan_2013}. Notably, the errors taken from \cite{McQuillan_2013} are very low and probably over-optimistic. The errors on $T_{eff}$ were taken from \cite{Berger_2020} for the samples from \cite{kamai2024} catalog. For samples from \cite{McQuillan_2013} catalog, no error was provided, so we added an arbitrary error of $100$ K. 

One of the characteristics of circumbinary systems is stability regions - planets can only be dynamically stable in specific regions. Stability is given by the ratio of the semi-major axis, which can be converted to the ratio of periods using Kepler's third law, and accounting for the stellar masses. We used stability criteria from \cite{Holman1999}, which provides stability tables for different eccentricities and masses. \cite{Holman1999} provides two tables - one for the case of an inner (circumstellar) planet and one for the case of an outer (circumbinary) planet. We took the minimum value of the outer table and the maximum value of the inner table as lower and upper bounds. If the system is indeed circumbinary, the period ratio must be above the minimal outer value, or below the maximal inner value, to be a binary-planet system with a planet around one star. Every planet found between those values should be unstable. 

\begin{figure*}
    \centering
    \includegraphics[width=0.6\textwidth]{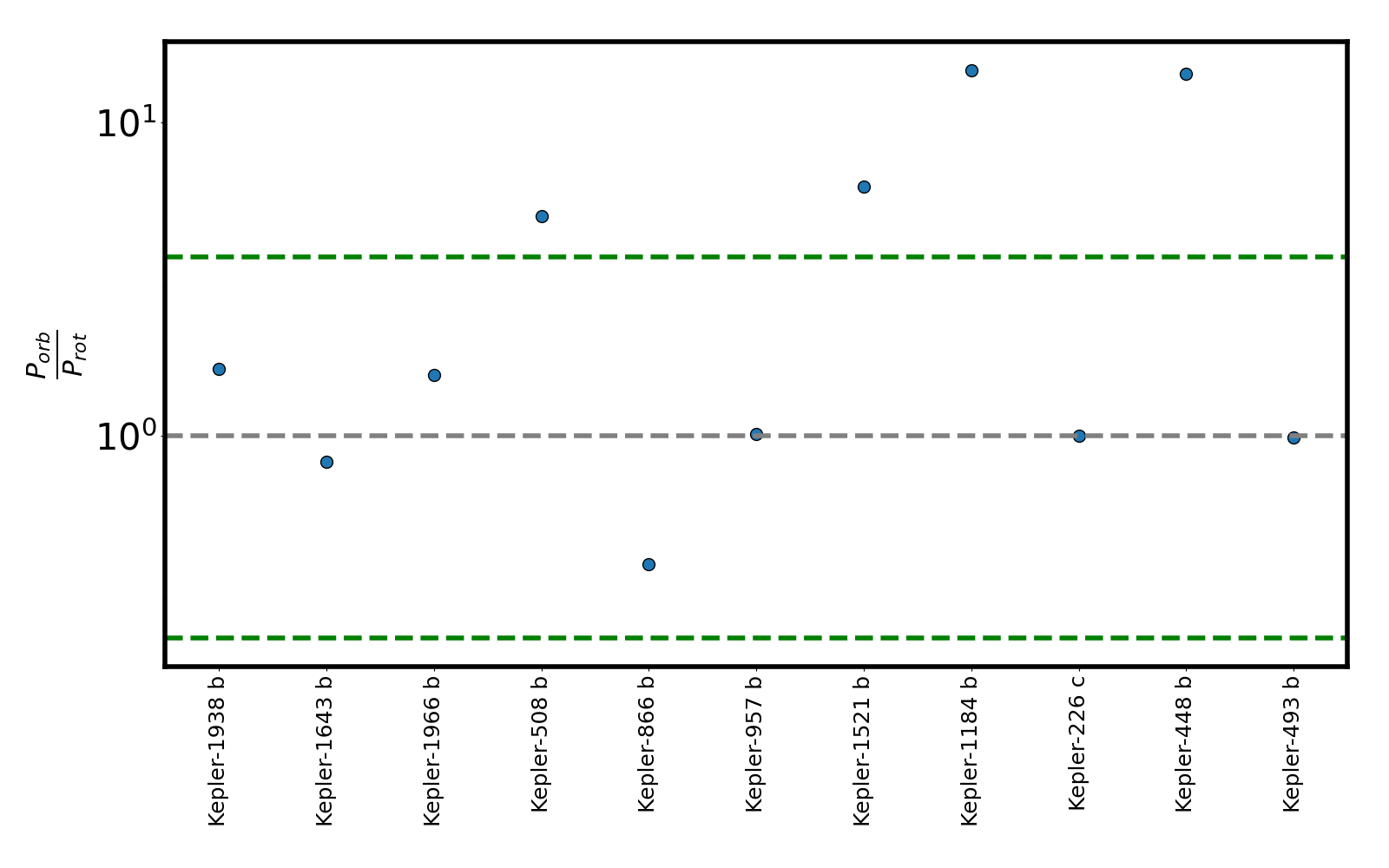}
    \caption{period ratio of the systems from Figure \ref{fig:planet_scatter}. The green lines represent circumbinary stability lines - the upper line represents the lower stability bound for an outer planet, and the lower line represents the upper stability bound for an inner planet. Both stability criteria are taken from \cite{Holman1999}. The gray line represents period ratio of $1$} \label{fig:planet_stability}
\end{figure*}

Figure \ref{fig:planet_stability} shows a scatter plot of the period ratios of the potential circumbinary systems. The dashed green lines represent stability bounds - the upper line represents the minimal value for an outer planet, and the lower line represents the maximal value for an inner planet. As mentioned before, for a system to be stable, the period ratio must be either above the upper line or below the lower line. Interestingly, we see that there are $7$ systems in the unstable region. We now investigate separately the stable and unstable systems.

\subsubsection{Potential Circumbinary planet systems} \label{subsec:circumbinaries}
\begin{figure*}
    \centering
    \includegraphics[width=0.6\textwidth]{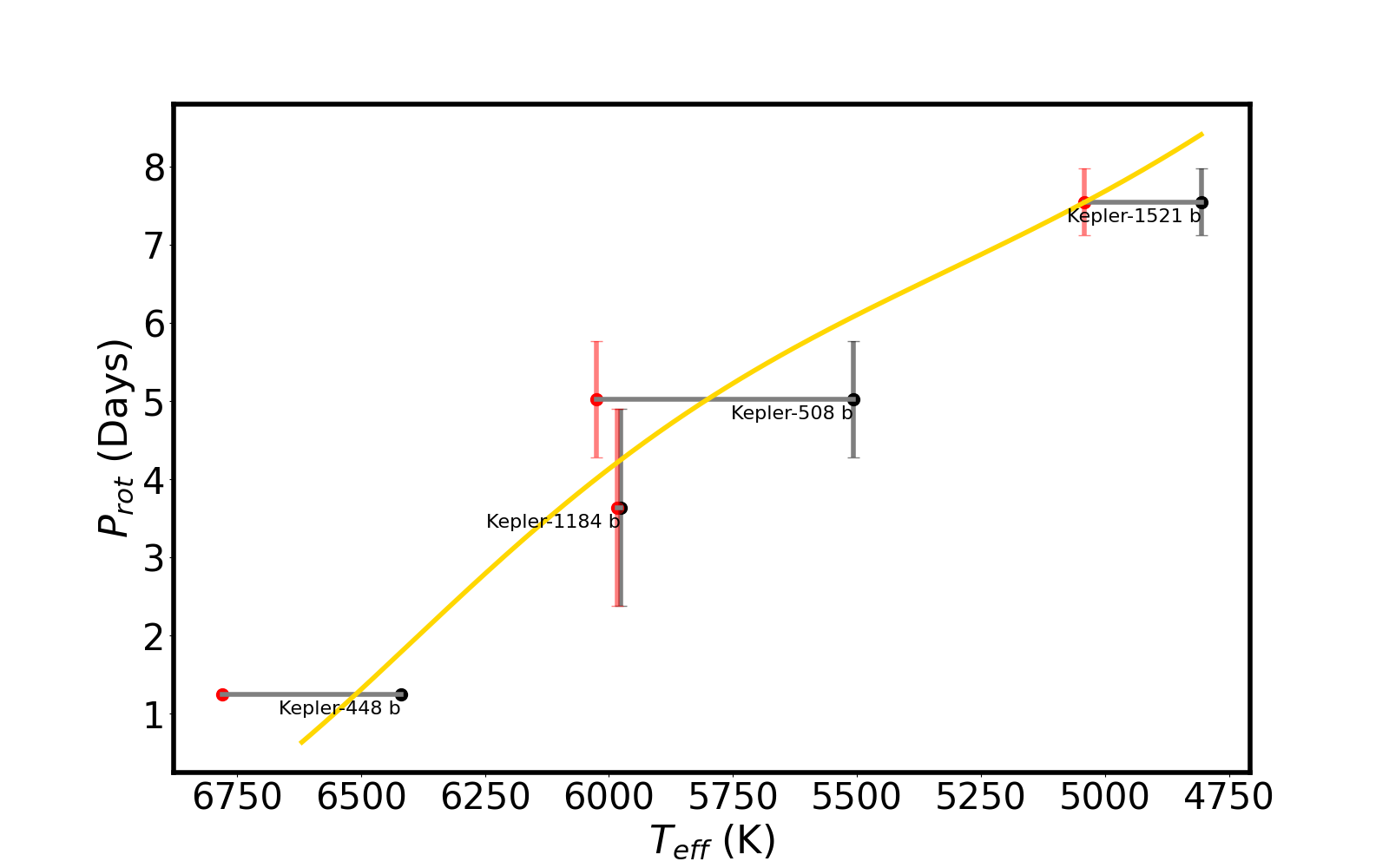}
    \caption{The rotation period as a function of $T_{eff}$ for circumbinary candidates (systems above the upper green line in Figure \ref{fig:planet_stability}). The black points correspond to $T_{eff}$ from \cite{Berger_2020} and the red points correspond to $T_{eff}$ from \cite{Morton_2016}. The period errors are taken from \cite{kamai2024} and \cite{McQuillan_2013}.} \label{fig:circum_teff}
\end{figure*}

There are $4$ systems above the stability line - \textit{Kepler 448 b}, \textit{Kepler 508 b}, \textit{Kepler 1184 b} and \textit{Kepler 1521 b}. First, we noted that their effective temperature can have large discrepancies between different resources. We used the temperature from \cite{Berger_2020}, but all four systems also have temperature measurements from \cite{Morton_2016}. To illustrate the effect of a discrepancy between temperature measurements, we show in Figure \ref{fig:circum_teff}, for each of the four systems, both temperature measurements (\cite{Berger_2020} in black and \cite{Morton_2016} in red). We see that for two systems, \textit{Kepler 448 b} and \textit{Kepler 508 b}, using the temperature from \cite{Morton_2016} moves the systems above the critical separation line, even within the period errors, i.e. their hosts are possibly more massive stars which could be fast-spinning even without being in a synchronized binary. Regarding \textit{Kepler 1521 b}- using the temperature from \cite{Morton_2016}, it moves to be exactly on the line. In addition, it was reported in \cite{Bouma2024} as a young system ($\sim 200$Myr), using both gyro and Lithium. We therefore also remove \textit{Kepler 1521 b} and treat only \textit{Kepler 1184 b} as a potential circumbinary system. We note that it is hard to verify if \textit{Kepler 1184 b} is indeed non-single star based on RUWE ($1.03$), $\Delta K_{iso}$ ($-0.63$), and $P(nss|T,P)$ ($0.72$) alone. The final classification requires a more detailed and dedicated investigation, which is beyond the scope of this paper.

\subsubsection{Planet synchronization and false positives}
\label{subsec:false_positives}
There are $7$ systems that are not stable according to \cite{Holman1999} criteria. Interestingly, most of them show a period ratio very close to unity, which suggests synchronization between the companion and the primary. In turn, this suggests one of the following possibilities:
\begin{itemize}
    \item The star is synchronized by the planet.
    \item The star is synchronized by a stellar companion that was misidentified as a planet (False Positive).
\end{itemize}

\begin{figure*}
    \centering
    \includegraphics[width=0.6\textwidth]{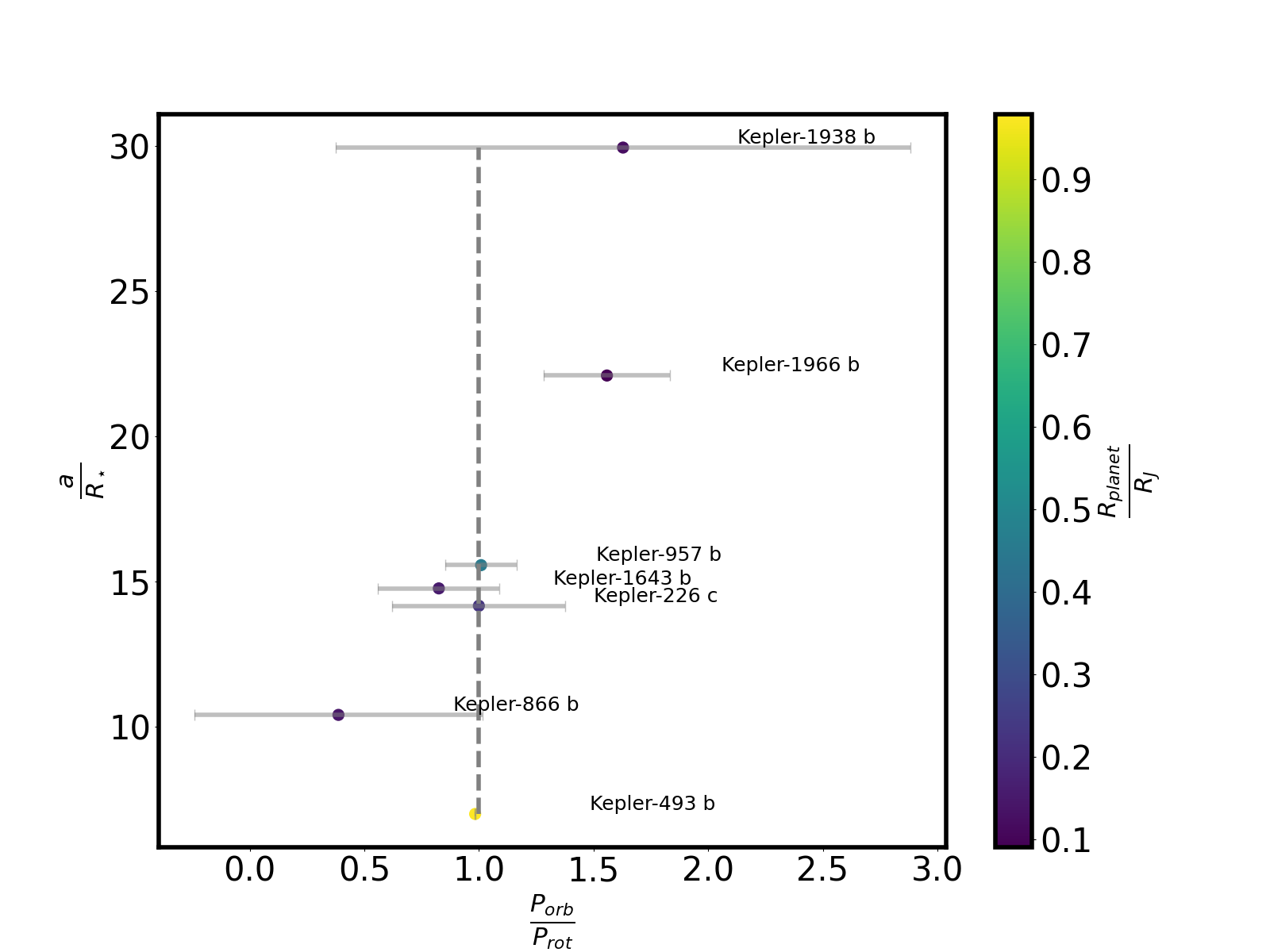}
    \caption{Period ratio as a function of $\frac{a}{R_\star}$ for systems below the upper green line in Figure \ref{fig:planet_stability}. Colors represent planet radius in units of Jupiter radii. } \label{fig:planet_fp}
\end{figure*}

For the first option, we need an efficient tidal dissipation process. Tidal torque strongly depends on $a/R_\star$ and $q$. Here, $a$ is the semi-major axis, $R_{\star}$ is the stellar radius, and $q$ is the mass ratio. For example, \cite{Knudstrup2024} analyzed 205 exoplanets and required  $\frac{a}{R_\star} < 10$ and $M_p > 0.3M_J$, where $M_J$ is Jupiter mass, for tidal obliquity alignment. In Figure \ref{fig:planet_fp}, we show the period ratio as a function of $a/{R_\star}$ for the $7$ non-stable systems. The colors represent the radius of the planet in units of Jupiter-radius. We see that one system, \textit{Kepler-493 b}, meets the criteria set by \cite{Knudstrup2024}. Another system, \textit{Kepler-957 b} has $a/{R_\star} = 15.6$ and ${R_p}/{R_J}=0.46$. Both systems have a period ratio of almost exactly unity ($0.98$ and $1.01$). In addition, \textit{Kepler-957 b} was reported by \cite{Bouma2024} as young (using only gyro), which reduces its probability of being a binary system. All other planets have much smaller inferred radii. Out of the remaining sample, two systems were reported as young ($\leq 200$ Myr) - \textit{Kepler-1938 b} (only gyro), and \textit{Kepler-1643 b} (gyro and Lithium), and were removed from the sample. The remaining three systems are potential false positives - misidentified stellar companions (e.g. grazing eclipses, misidentified as planet transits). We conclude that among the $7$ systems, two are possibly synchronized by a planet, two are possibly young systems, and $3$ are likely false-positive stellar binary systems. Follow-up radial-velocity measurements, if possible, should, in principle, identify if these are indeed stellar binaries.

\subsubsection{Planets orbiting ultra-fast spinning stars}
We finish this section by discussing potential planets orbiting very fast rotators ($P_{orb} < 3$), which are potential triple systems. It is suggested that it is unlikely to find non-massive planets around a short-period binary \citep{Welsh2015, Armstrong2014, Hamers2016}. The reason is that massive planets can affect the process of Kozai-Lidov cycles and tidal friction (KLCTF), which is possibly responsible for the creation of short-period binaries. Non-massive planets would not affect KLCTF, but this implies one of the following options - either the planet is unstable, or it is sufficiently wide and has a high inclination with respect to the inner binary (for KLCTF to be possible). Given the high mutual inclination, the last two options imply that such a planet would likely not be identified through a transit around an eclipsing binary. In principle a planet transits a binary at a relatively short period, it might be possible to detect transit-time variations (TTV), which will show the existence of a companion inner star. In any case, we found none of the candidates to have detected TTVs. We also looked for a cross-match between confirmed planet host stars and our sample of potential triple systems, described in \ref{subsec:triples}. We find only one such system - \textit{Kepler-1644}. Further investigation reveals that the system is probably a false positive \citep{Wang2024} (note that they refer to the KIC identification, not the Kepler planet naming). To conclude, we didn't find any confirmed planet orbiting a very short period binary, consistent with the suggestion of \cite{Hamers2016}.

\section{Tidal features in known binaries} \label{sec:tidal_features}
We now move from identification to investigation of known non-single stars. Specifically, we investigate tidal effects. \\
Tidal synchronization involves three long-term processes: 
\begin{itemize}
    \item Synchronization of the stellar and orbital period.
    \item Alignment of stellar rotation axes with orbital rotation axis. 
    \item Circularization of the orbit.
\end{itemize}

\begin{figure*}
    \centering
    \begin{minipage}[b]{0.45\textwidth}
    \includegraphics[width=1.1\textwidth]{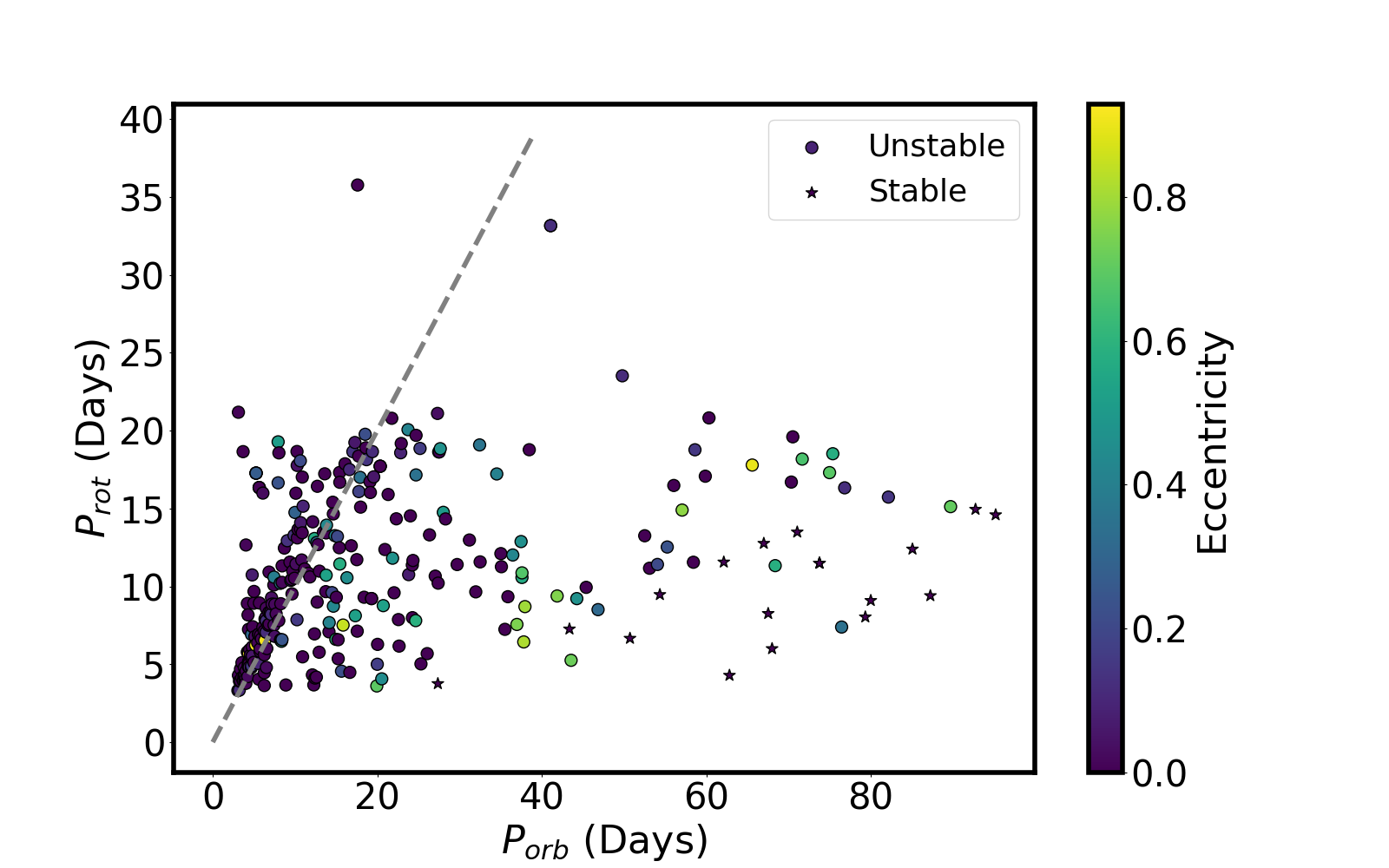}
    \end{minipage}
    \begin{minipage}[b]{0.45\textwidth}
    \includegraphics[width=1.1\textwidth]{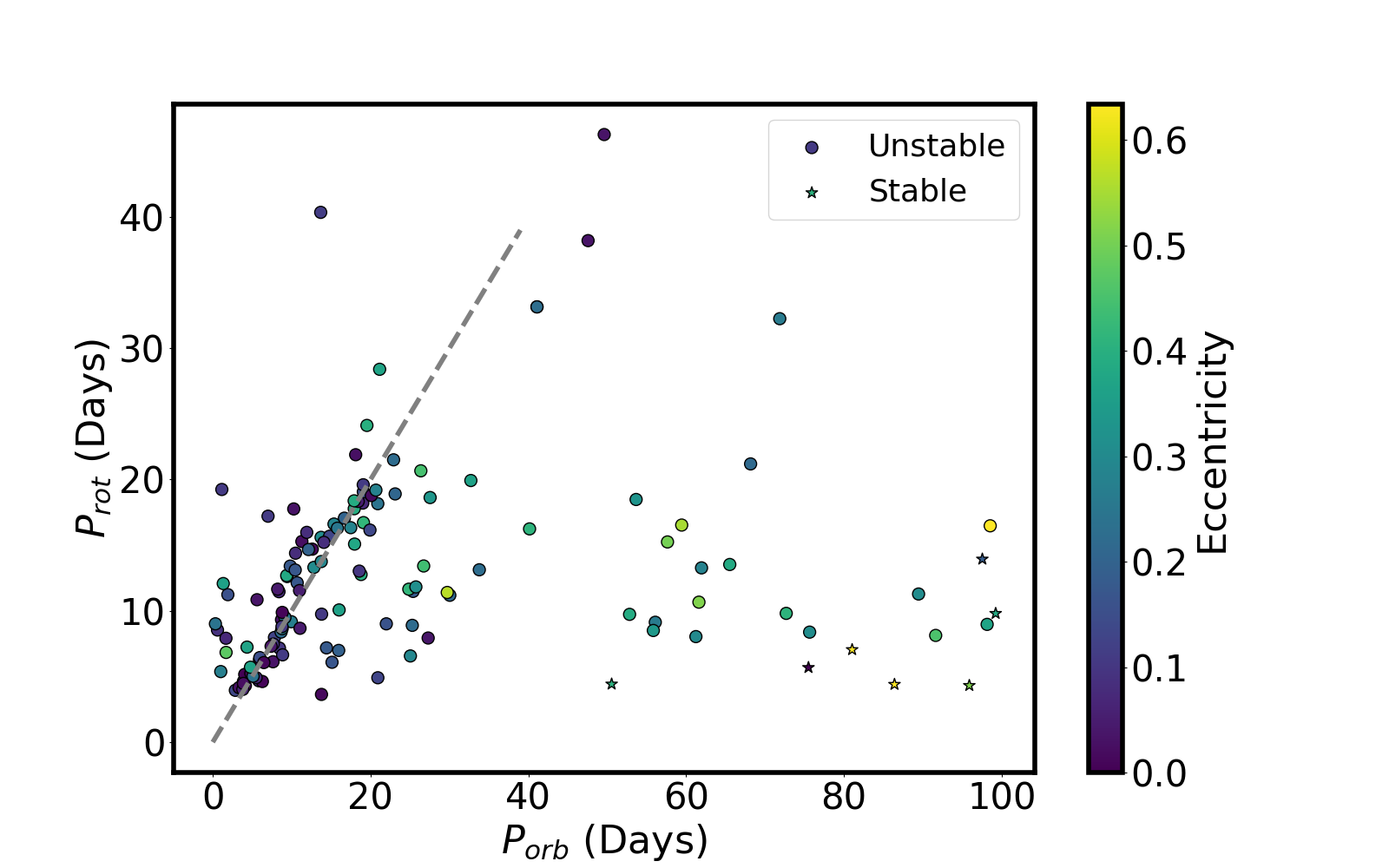}
    \end{minipage}
    \caption{Orbital period vs stellar period. The left panel shows eclipsing binaries, and the right panel shows Gaia non-single stars. The dashed line represents a line with a slope of unity. } \label{fig:tidal_sync_p}
\end{figure*}

\begin{figure*}
    \centering
    \begin{minipage}[b]{0.45\textwidth}
    \includegraphics[width=1.1\textwidth]{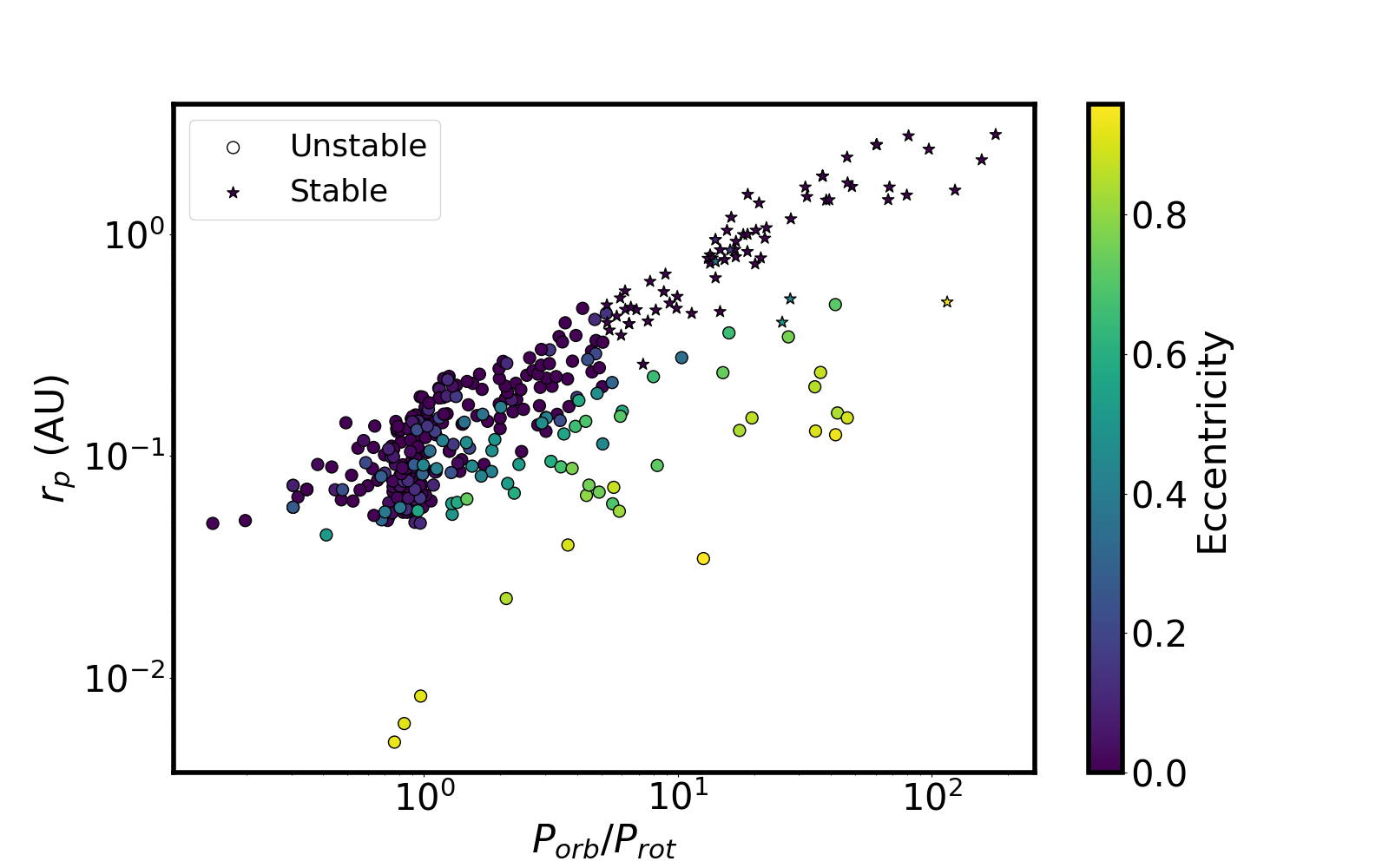}
    \end{minipage}
    \begin{minipage}[b]{0.45\textwidth}
    \includegraphics[width=1.1\textwidth]{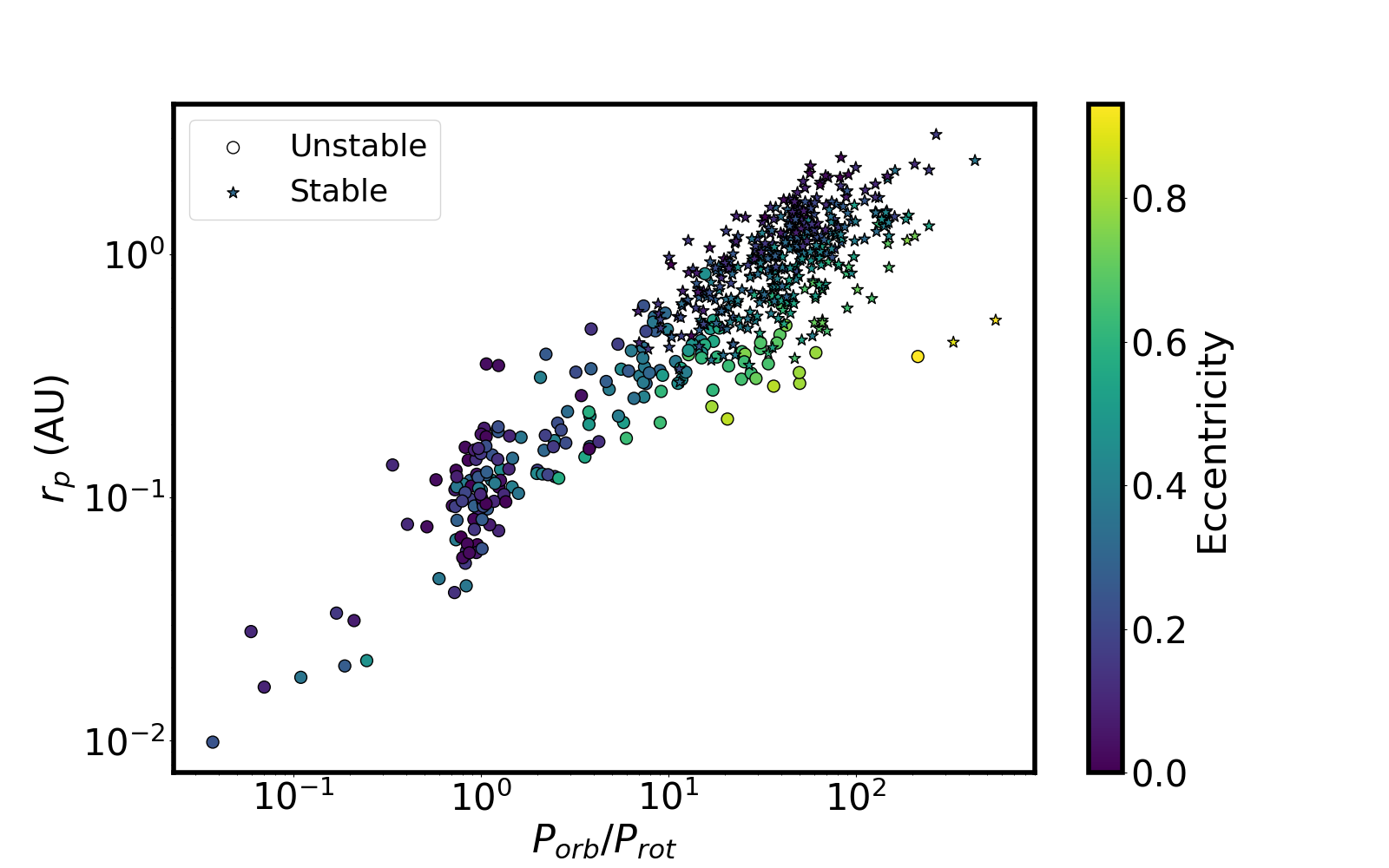}
    \end{minipage}
    \caption{Period ratio vs pericenter distance. Color represents eccentricity. The left panel shows \textit{Kepler} eclipsing binaries, and the right panel shows Gaia-nss samples. Star symbols correspond to systems that could be stable as triples, taking the orbital period as the outer orbit with the inner orbit being the synchronized binary. See the text for details.} \label{fig:tidal_r_p}
\end{figure*}

\begin{figure*}
    \centering
    \begin{minipage}[b]{0.45\textwidth}
    \includegraphics[width=\textwidth]{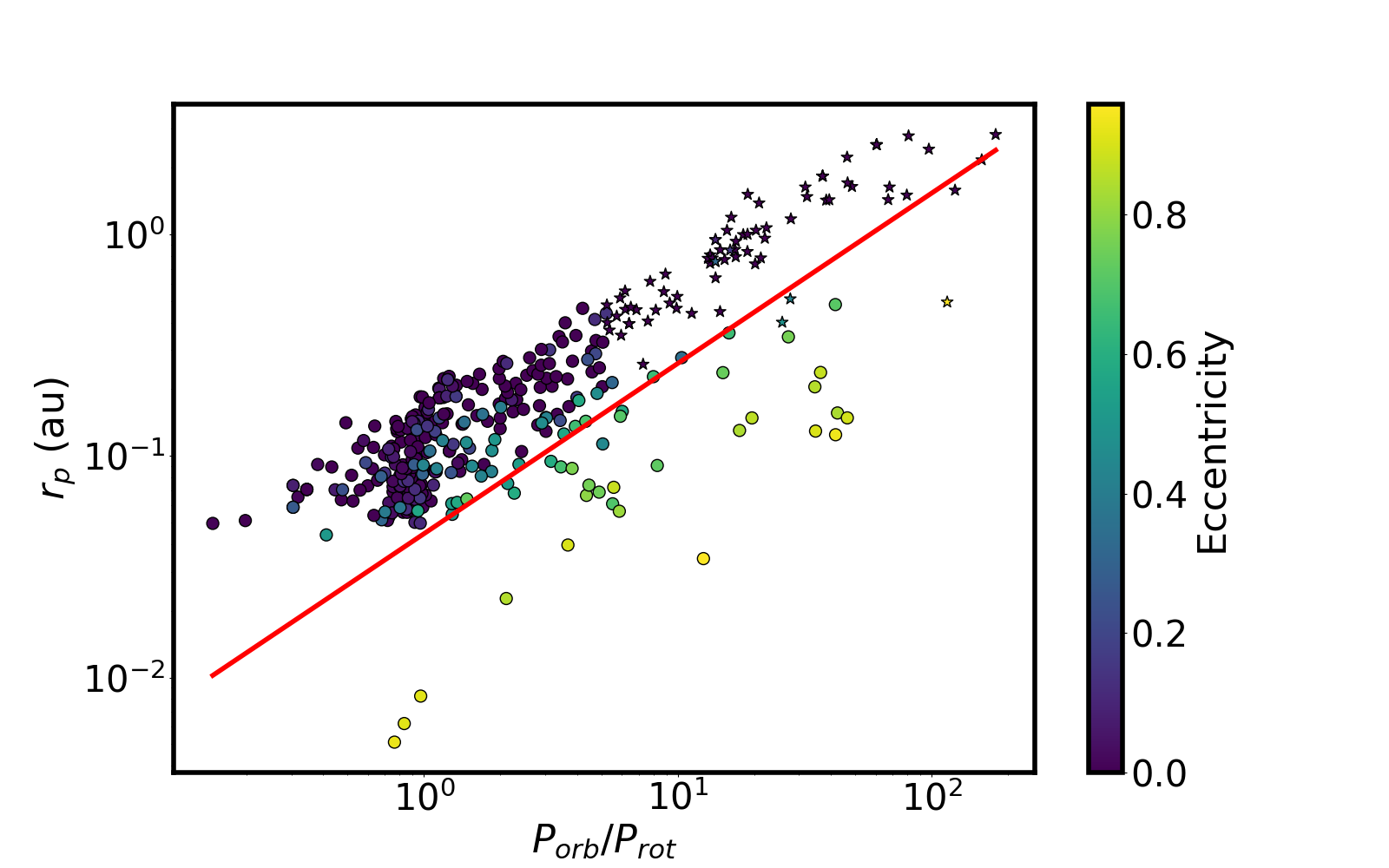}
    \end{minipage}
    \begin{minipage}[b]{0.45\textwidth}
     \includegraphics[width=\textwidth]{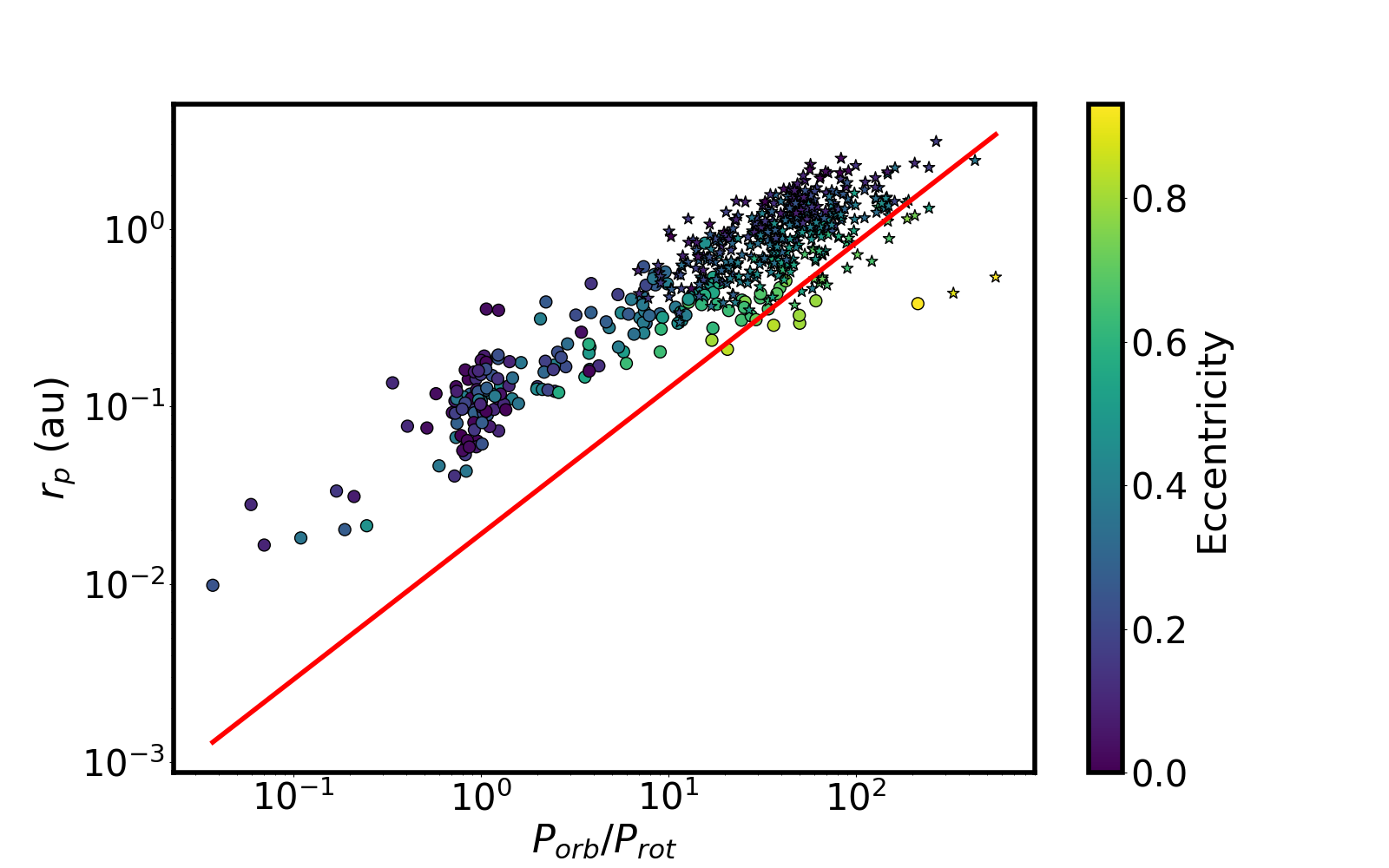}
    \end{minipage}
    \caption{Period ratio vs pericenter distance with the best fit of the lower envelope according to equation \ref{eq:envelope_fit}. Colors represent eccentricity. The left panel shows \textit{Kepler} eclipsing binaries, and the right panel shows Gaia-nss samples.  } \label{fig:tidal_r_p_fit}
\end{figure*}

We note that these processes have different time scales. The reason is that the stellar angular momentum is much smaller than the orbital angular momentum. As such, much less angular momentum is needed to synchronize the stellar period and rotation axes than to circularize the orbital period. Therefore, we expect that period synchronization and alignment would occur on much shorter timescales than circularization and that all circularized orbits would be synchronized and aligned. For a detailed review of tidal effects, see \cite{Mazeh2008} and \cite{Zahn2008}. \\
We want to see if we can observe tidal effects in known binaries using periods from \cite{kamai2024}. 
The EBs sample was already analyzed by \cite{Lurie2017} and \cite{kamai2024}. However, they didn't make use of measured eccentricities. Here, we use EB eccentricities, calculated by \cite{IJspeert2024}. In addition, we also analyze the Gaia-nss sample, which used different methods for binarity identification and provides eccentricities as part of the orbital solutions. While \cite{Bashi2023} studied the tidal features of Gaia-nss catalog for general stars, they didn't use information about the stellar rotation period, which we use. This combination of two independent datasets should increase the robustness of our results. We cross-matched Gaia-nss and EBs with the catalog from \cite{kamai2024} which resulted in $575$ Gaia-nss samples and $378$ EBs samples. Figure \ref{fig:tidal_sync_p} shows the orbital period vs. the stellar period for EBs (left panel) and Gaia-nss (right panel). The graphs were cut at $P_{orb} = 100$ days for visual purposes. In both panels, the dashed line represents the period synchronization line, the colors represent eccentricities, and the shape of the points represents stability as a hierarchical triple system. We further explain the last point; to check the possibility that the identified binary is a triple system with an inner unidentified binary and outer companion, we checked the stability criteria for such systems using the algebraic criteria given in \cite{Vynatheya2022}. We used the identified orbital period as the outer period and the stellar period as the synchronized period of an inner binary. Since we assumed synchronization for the inner binary, we used zero inner eccentricity and took the measured eccentricity as the outer one. We assumed an equal mass ratio among all three companions and zero inclination between the outer companion and the inner binary. Using these simplified assumptions, we were able to calculate the stability for each sample using equation 4 in \cite{Vynatheya2022}. In Figure \ref{fig:tidal_sync_p}, points with circular shape are unstable as a hierarchical triple system, and star shaped symbols are stable.

We can see that both samples show period synchronization up to $\sim 20$ days. This aligned with the finding of \cite{Fleming2019}. They simulated two different models of equilibrium tidal torque together with a magnetic braking model, and analysed the competition between the two. They found that according to the constant time lag (CTL) model, the probability of a binary to be tidally synchronized drops dramatically for $P_{orb} > 20$ days, which implies that magnetic braking stars overpower tidal forces at this regime. On the other hand, the constant phase lag (CPL) model predicts synchronization at much larger $P_{orb}$. The fact that, on both samples, we see synchronization up to 20 days and almost no evidence for synchronization for longer periods strongly suggests that the CTL model is preferred. We also see that in both panels, most of the synchronized samples are indeed circular, as expected. Looking at non-synchronized samples, we again see the same phenomenon in both panels: a group of eccentric, sub-synchronous samples exists up to an orbital period of $~40$ days. In longer orbits, we start to see the possibility of triple systems as stable configurations appear. This explains the existence of long orbital periods combined with fast stellar periods. In fact, most of the samples with $P_{rot} < 10$ days are either synchronized, in the process of synchronization, or might be a synchronized binary in a hierarchical triple system. This is in agreement with the finding of \cite{Lurie2017}, that most binaries with orbital periods $< 10$ days are synchronized and circular. Both panels suggest a separation between the populations of non-synchronized binaries and possible triple systems at $~40$ days. 

We can use the period and eccentricity to calculate the pericenter, $r_p$, which is the closest point in an eccentric orbital period. We assume equal masses and use the mass from \cite{Berger_2020}. We then estimated the semi-major axis using Kepler's third law. Then, using the eccentricity, we can calculate the pericenter using 
\begin{align}
    r_p = a(1-e) ;
\end{align}
Where $a$ is the semi-major axis and $e$ is the eccentricity.
Figure \ref{fig:tidal_r_p} shows the period ratio $\frac{P_{orb}}{P_{rot}}$ vs. the pericenter distance for EBs (left) and Gaia-nss (right). Colors represent eccentricity, and the shapes of the points represent stability in the same way as in Figure \ref{fig:tidal_sync_p}. We recognize a difference between the Gaia-nss sample and the EBs sample on low period ratios and low $r_p$ - Gaia-nss samples show a curved shape in the envelope in this regime. This pattern is not seen in EBs samples and can be understood as 'missing' samples due to detectability limits of Gaia-nss samples as discussed in section \ref{sec:data}. Regardless of their difference, both panels show a clear synchronization envelope, which sets a lower bound on $r_p$. This lower bound can be seen as a lower bound on the mass ratio, since our assumption of equal mass is not true. Similarly to Figure \ref{fig:tidal_sync_p}, both panels show a group of sub-synchronous eccentric orbits close to the envelope, and a total of four samples (three in Gaia-nss and one in EBs) with very large periods ratio ($\frac{P_{orb}}{P_{rot}} > 100$) and very small pericenter ($r_p < 0.6$ AU). It is known that high eccentricities can result from a hierarchical system with a distant third companion \citep{vonZeipel1910, Lidov1962, Kozai1962}
Interestingly, three out of those four samples are triple-stable. This suggests further investigation, which is beyond the scope of this paper. We summarize the parameters of those four samples in Table \ref{table:r_p outliers}. 

We can better characterize the dependency of the period ratio with $r_p$ by fitting the lower envelope with a function. We use:
\begin{align} \label{eq:envelope_fit}
    f(P) = \alpha (\frac{P_{orb}}{P_{rot}}) ^ \beta
\end{align}
To fit only the lower envelope, we divided the values of $\frac{P_{orb}}{P_{rot}}$ into 40 bins and chose the minimum $r_p$ at each bin. Then, we fitted Eq. \ref{eq:envelope_fit} to the set of bins and minimal values. Figure \ref{fig:tidal_r_p_fit} shows the best fits for EBs (left) and Gaia-nss (right). The best-fit values are $\alpha_{gaia}=0.022$, $\beta_{gaia}=0.772$, $\alpha_{EBs}=0.045$, $\beta_{EBs}=0.769$. In both cases, we get $\frac{P_{orb}}{P_{rot}} \propto r_p^{\sim 0.77}$. The very good agreement in fitted power law between two independent datasets suggests that this is a physical phenomenon that describes the evolution of $r_p$ during the synchronization process and not effects related to a specific dataset.  

\section{Discussion and Summary} \label{sec:conclusions}
In this study, we have developed and applied a novel yet simple method to identify non-single stars using rotation period and temperature only. By establishing a separation line in the period-temperature space, we identified $2229$ potential non-single stars in the \textit{Kepler} field, providing a robust tool for distinguishing single stars from binaries and higher-multiplicity systems. Using analysis of the distributions of the peculiar velocities of non-single candidates, we were able to identify non-singles and separate them from the young population, using $\sigma = 12 \, km \cdot s^{-1}$ as a cutoff velocity. For stars with no known $\sigma$, we assess the probability of each candidate to be a non-single star and not a young star. The average probability was found to be $72\%$, demonstrating the effectiveness of this method.

Given that stars with rotation periods below three days are synchronized binaries, as discussed above, and that binaries with such short periods are likely to reside in triple systems, we photoemetrically identify 1518 triple star candidates in the Kepler field. This is the largest catalog of triple-star candidates homogeneously identified (e.g. compare with \cite[][and references therein]{Cza+23,Bas+24,Bor+25}. 

Applying our method to planet-host stars, we identified a potential circumbinary system (\textit{Kepler 1184} and two systems possibly synchronized by close-in gas-giant planets (\textit{Kepler 493} and \textit{Kepler 957}); alternatively, these systems could be grazing eclipsing binaries misidentified as planets. We also argued that three other systems are likely false positives, with the detected "planets" being stellar companions. These findings underscore the importance of considering binarity in exoplanet studies, as misidentified stellar companions can significantly impact our understanding of planetary systems.

These results open the door for further discoveries using upcoming or existing photometric datasets. Future applications of our method to surveys such as TESS and PLATO could uncover additional non-single systems and help refine our understanding of stellar multiplicity across different stellar populations and environments.

Our analysis of tidal features in known non-single stars revealed clear evidence of period synchronization, orbit circularization, and constraints on the minimal pericenter radius during the synchronization process. We construct a phenomenological characterization of this process using the relation $r_p \propto (\frac{P_{orb}}{P_{rot}})^{0.77}$, a result consistent across two independent datasets (Eclipsing Binaries and Gaia non-single stars). Additionally, we identified three systems as potential hierarchical triples with high eccentricities, highlighting the complex dynamics of multi-star systems.

The results presented here have broad implications for stellar and planetary astrophysics. By providing a clearer picture of tidal evolution and binarity, our work contributes to a deeper understanding of stellar angular momentum evolution, binary formation, and the stability of planetary systems in binary environments.

\section{acknowledgments}
 We would like to acknowledge the support from the Minerva Center for life under extreme planetary conditions.

\begin{table*}
\centering
\begin{tabular}{||c c p{2.5cm} p{2.5cm} c||}
 \hline
Kepler ID & Gaia ID             & Orbital Period (Days) & Stellar Period (Days) & Eccentricity \\
 \hline
5217805   & 2073546374433598592 & 1668.97               & 5.04                  & 0.878        \\
 \hline
6933899   & 2105473993062697856 & 4063.67               & 7.30                  & 0.917        \\
 \hline
8424629   & 2126925842879471744 & 2754.99               & 12.87                 & 0.931  \\
\hline
9408440 & - & 989.98 & 8.61 & 0.808 \\
 \hline
 \hline
\end{tabular}
\caption{Parameters of binaries with $\frac{P_{orb}}{P_{rot}} > 100$ and $r_p < 0.6$ AU. See Figure \ref{fig:tidal_r_p}.} \label{table:r_p outliers}
\end{table*}

\bibliographystyle{aasjournal}
\bibliography{main}

%\appendix

\end{document}